\documentclass[%
 reprint,
 amsmath,amssymb,
 aps, sort&compress
]{revtex4-2}
\usepackage{xcolor}
\usepackage{graphicx}% Include figure files
\usepackage{dcolumn}% Align table columns on decimal point
\usepackage{bm}% bold math
\usepackage{ragged2e} %% used for justification purpose. You can use now `\begin{justify}...\end{justify}` environment.
\usepackage[hidelinks]{hyperref}% add hypertext capabilities without ugly boxes around the citations.
\hypersetup{
  colorlinks   = true, %Colours links instead of ugly boxes
  urlcolor     = blue, %Colour for external hyperlinks
  linkcolor    = blue, %Colour of internal links
  citecolor   = blue %Colour of citations
}
\begin{document}

\preprint{APS/123-QED}

%\title{Intercellular Adhesion Drives Fluid To Solid Transition In A Model Tissue }

\title{Role of intercellular adhesion in modulating tissue fluidity}

%\\with Forced Linebreak}% Force line breaks with \\

\author{Soumyadipta Ray$^1$}
 \author{Santidan Biswas$^2$}%
 \email{sab231@pitt.edu}
\author{Dipjyoti Das$^1$}%
 \email{dipjyoti.das@iiserkol.ac.in}
\affiliation{%
 $^{1}$Department of Biological Sciences, Indian Institute of Science Education And Research Kolkata, Mohanpur, Nadia - 741 246, West Bengal, India \\
 $^2$Chemical Engineering Department, University of Pittsburgh, Pittsburgh, PA 15261, USA %\textbackslash\textbackslash
}%



%\affiliation{%


%\author{Dipjyoti Das}
%\affiliation{
%\affiliation{
 %Third institution, the second for Charlie Author
%}%
%\author{Dipjyoti Das}
%\affiliation{%



\begin{abstract}

Tuning cell rearrangements is essential in collective cell movement that underlies cancer progression, wound repair, and embryonic development. A key question is how tissue material properties and morphology emerge from cellular factors such as cell-cell adhesion.  Here, we introduce a two-dimensional active force-based model of tissue monolayers that captures the liquid-to-solid transition exhibited by tissues. Unlike the Vertex and Voronoi models, our model shows that reducing intercellular adhesion in near-confluent tissues leads to spontaneous neighbor exchanges and fluidization.   Near the liquid-solid phase boundary, we also found glassy behavior characterized by subdiffusive dynamics, swirling cell motion, and non-Gaussian exponential tails in displacement distributions. These exponential tails collapse onto a single master curve, suggesting a universal 'diffusion length' in the glassy regime. Notably, we demonstrate that structural parameters based on cell shape cannot always distinguish tissue phases due to huge cell shape fluctuations that are not observed in Vertex and Voronoi models. Our general simulation framework streamlines previous approaches by removing many arbitrary features and can reproduce known model behaviors under different conditions, offering potential applications in developmental biology and physiology.

%\begin{description}
%\end{description}
\end{abstract}

\maketitle

%\tableofcontents

\section{\label{sec:level1} INTRODUCTION } 

%{\bf INTRODUCTION:}
Tissue fluidity is modulated in diverse biological contexts such as embryonic development \cite{Mongera2018, pinheiro2022morphogen}, wound healing \cite{poujade2007collective, vishwakarma2020dynamic}, and cancer progression \cite{Kim2020, oswald2017jamming}. A solid-to-fluid transition can sculpt tissues during tail elongation in developing zebrafish embryos \cite{Mongera2018, Kim2021, pinheiro2022morphogen}. This phase transition also coincides with the cell-state switching and changes in gene expressions, called Epithelial-to-Mesenchymal transition (EMT) \cite{Das2017, Mongera2018}. EMT has been proposed to mimic an unjamming transition in tissues \cite{Sadati2013, Park2015, Mitchel2020}. Fluid-solid transitions were also observed in cell monolayers \cite{Garcia2015, Angelini2011}, where transitions occur either with increasing cell density (due to proliferation) \cite{Angelini2011} or with increasing cellular adhesion \cite{Garcia2015}. Tissue monolayers further undergo solid-to-fluid transitions by applying external compressive stress \cite{Park2015, Mitchel2020}. How tissue-level transitions emerge from governing cellular parameters is an important question yet to be fully understood \cite{atia2021cell, Sadati2013, trepat2018mesoscale}.

Two classes of computational frameworks mainly exist to explain tissue-level transitions: Self-propelled particle (SPP) models and models that treat cells as extended objects. SPP models describe cells as soft discs \cite{szabo2006, Belmonte2008, sepulveda2013collective} and have been used to study the density (or packing fraction) dependent \cite{Henkes2011, Fily2014,fily2012athermal, Berthier2014, ni2013pushing} and motility-driven \cite{Henkes2011, Fily2014, ni2013pushing, Garcia2015} jamming transitions. However, SPP models can not capture the cell shape changes in a confluent tissue where there are almost no gaps between cells (i.e., packing fraction close to unity). Meanwhile, the Vertex model and its variants \cite{nagai2001dynamic, Farhadifar2007, Bi2015, bi2014energy, hufnagel2007mechanism, li2019mechanical, yan2019multicellular, Das2021}, Self-propelled Voronoi model \cite{Bi2016, li2014coherent, barton2017active}, Deformable Particle model \cite{dp2018prl}, Cellular Potts model \cite{szabo2010collective, Chiang2016, sadhukhan2021theory}, and Phase Field model \cite{loewe2020solid} describe cells as extended objects accounting for the cell geometry. Cell dynamics in all these models depend on energy functions, unlike SPP models that describe the force-based motion of cell centers. For instance, 2D Vertex and Self-propelled Voronoi (SPV) models define a shape-energy function that depends on each cell's target area ($a_0$) and perimeter ($p_0$). These models predict a density-independent solid-to-fluid transition with an increasing target shape index, $s_0$, defined as a dimensionless parameter ($s_0=p_0/\sqrt{a_0}$). However, the increase in shape index was interpreted as an increase in cell-cell adhesion \cite{Bi2015, Bi2016}. This counter-intuitively implies that increasing adhesion drives tissue fluidization.

 In contrast, increasing cell-cell adhesion induces solidification in MDCK cell monolayers \cite{Garcia2015, choi2022cell} and organotypic cultures \cite{ilina2020cell}.
Also, the downregulation of adhesion molecules in the elongating zebrafish tail coincides with EMT and tissue fluidization \cite{Das2017, Mongera2018}.  Thus, the role of intercellular adhesion on the tissue scale transitions is still debatable.
We address this by introducing a 2D active force-based model of tissue monolayers, where each cell made with beads and springs represents a soft deformable object. 

We combine the particle-like nature and force-based dynamics of SPP models with the spatial extension of cells as in Vertex and SPV models, eliminating the deficiencies of both types of models. For instance, our model displays spontaneous neighbor exchanges, while Vertex and SPV models implement these events in an ad hoc manner, using a threshold value on cellular edge lengths. Additionally, the interface of adjoining cells in Vertex and SPV models is represented by a shared (common) edge, making it impossible for adjoining cells to move relative to one another, thereby dictating a geometric constraint \cite{manning2023essay}. Our model removes this unphysical constraint, allowing for spontaneous fluctuations in cell shapes. Notably, our model significantly differs from topological energy-based models, like the Vertex and Voronoi models, where a single cell loses its physical meaning in isolation (see Section III in SI).

In our model, increasing cell-cell adhesion leads to liquid-to-solid transition, contrary to the predictions of Vertex and Voronoi models. Our observation is consistent with {\it in vitro} experiments on tissues  \cite{Garcia2015, ilina2020cell} and {\it in vivo} experiments on morphogenesis during embryonic development \cite{Das2017, Mongera2018}. Near the liquid-solid phase boundary, we also observe glassy dynamics with many signatures of dynamic heterogeneity. 
In the glassy regime, we collapse distinct exponential tails of displacement distributions onto a master curve, revealing a universal `diffusion length' that explains the origin of dynamic heterogeneity. Our framework provides precise control over individual cell properties and cell-cell interaction to gain mechanistic insights into emerging tissue behavior, with potential applications in various developmental and physiological contexts.

\section{\label{sec:level1} MODEL }

\begin{figure}[hbt!]
\includegraphics[width=8.6cm]{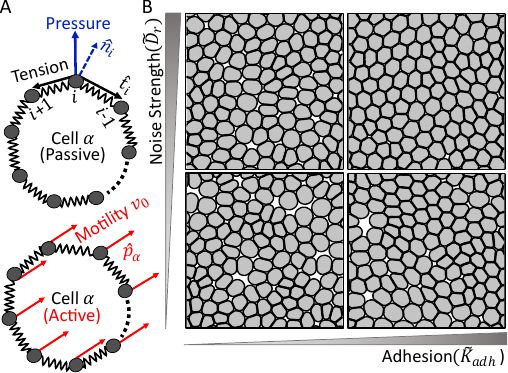}% Here is how to import EPS art
\caption{\label{fig:epsart} \textbf{Model and configurations}. \textbf{(A)} A single cell modeled as a closed loop of beads and springs. Each bead experiences an outward-normal pressure and tangential spring forces (Top). Additionally, in an active cell, all beads move with a self-propulsion speed $v_o$ along a noisy polarity direction, $\hat{p}_\alpha$ (Bottom). \textbf{(B)} Steady-state tissue configurations. Increasing adhesion and noise strength can lead to cell jamming with reduced intercellular space.}
\label{fig1}
\end{figure}

%{\bf MODEL:} 
We adopted a force-based passive mechanical model of 2D soft grains (or cells) \cite{Astrom2006pre, Mkrtchyan2014, Madhikar2021} instead of commonly used frameworks of energy-based Vertex or Voronoi models. A single cell is assumed to be a closed loop of beads connected by springs with stiffness $K_s$ and natural length $l_0$ (Fig.~\ref{fig1}A(top), Fig. S1). Each bead encounters tangential tension from adjacent springs and internal pressure, $P$. These intracellular forces capture the roles of actomyosin cortex and cytoplasmic pressure, respectively. Thus, the total intracellular force experienced by the $i$-th bead in a cell is:

\resizebox{0.9
\linewidth}{!}{
\begin{minipage}{\linewidth}
\begin{align}
\boldsymbol{F}_i^{intra} =\;\;& {K_s}({l_i}-{l_0})\hat{t}_i - {K_s}({l_{i+1}}-{l_0})\hat{t}_{i+1} + \frac{P{l_0}}{2}(\hat{n}_i + \hat{n}_{i+1}).
\label{eq:1}
\end{align}
\end{minipage}}

Here $l_i$ is the bond length between $i$-th and $(i-1)$-th beads; while $\hat{t}_i$ and $\hat{n}_i$ denote the tangential and the outward normal unit vectors corresponding to the $i$-th bead, respectively (Fig.\;S1). 
Moreover, beads of distinct cells experience short-ranged intercellular forces consisting of two parts: a spring-like attraction, describing the cell-cell adhesion, and a spring-like repulsion, preventing cell interpenetration. Thus, intercellular forces on the $i$-th bead of one cell due to the interaction with the $j$-th bead of another cell are
\resizebox{0.95
\linewidth}{!}{
\begin{minipage}{\linewidth}
\begin{align}
\boldsymbol{F}^{inter}_{ij} &= K_{adh}({r_c}^{adh}-r_{ij})\hat{r}_{ij}\;\; , \text{if}\;\; {r_c}^{rep}\leq r_{ij}\leq {r_c}^{adh} \nonumber\\
&=-K_{rep}({r_c}^{rep}-r_{ij})\hat{r}_{ij}\;\; , \text{if} \;\;r_{ij}< {r_c}^{rep} \nonumber\\
&=0 \;\;\; ,\;\;\; \text{otherwise}.
\label{eq:2}
\end{align}
\end{minipage}}

The distance between two interacting beads is $r_{ij} = |{\boldsymbol{r}_{i,\alpha}-\boldsymbol{r}_{j,\beta}}|$, and the corresponding unit vector is $\hat{r}_{ij} = (\boldsymbol{r}_{i,\alpha} - \boldsymbol{r}_{j,\beta})/r_{ij}$, where $i$-th and $j$-th beads belong to $\alpha$-th and $\beta$-th cell, respectively. Adhesive and repulsive forces are characterized by corresponding strengths ($K_{adh}$ and $K_{rep}$ respectively) and cut-off ranges (${r_c}^{adh}$ and ${r_c}^{rep}$).

% As in SPP \cite{Henkes2011} and SPV \cite{Bi2016} models, 

Additionally, we incorporate a polarity vector $\hat{p}_\alpha = (\cos\theta_\alpha, \sin\theta_\alpha)$, in which direction each bead of $\alpha$-th cell exerts a self-generated motility force of magnitude $cv_0$ (where $c$ is the coefficient of viscous drag and $v_0$ is the self-propulsion speed). All beads in each cell possess the same motility direction (Fig.~\ref{fig1}A(bottom)), but it differs from cell to cell. Together, the equation of motion governing the overdamped dynamics of the $i$-th bead of $\alpha$-th cell becomes
\resizebox{0.95
\linewidth}{!}{
\begin{minipage}{\linewidth}
\begin{align}
c\dot{\boldsymbol{r}}_{i,\alpha}=\boldsymbol{F}_{i,\alpha}^{intra} + \sum_{j}\boldsymbol{F}_{ij,\alpha}^{inter} + c{v_0}\hat{p}_\alpha.
\label{eq:3}
\end{align}
\end{minipage}}

Here, $j$-th bead belongs to another cell, interacting with the $\alpha$-th cell. The unit polarity vector for the $\alpha$-th cell further undergoes a rotational diffusion given by
\resizebox{0.95
\linewidth}{!}{
\begin{minipage}{\linewidth}
\begin{align}
\frac{\partial{\theta_{\alpha}}}{\partial{t}} &= \zeta_\alpha(t)\;\; ; \quad \langle\zeta_\alpha(t)\zeta_\beta(t')\rangle = 2D_r\delta(t-t')\delta_{\alpha\beta}. 
\label{eq:4}
\end{align}
\end{minipage}}

Here, the angle $\theta_{\alpha}$ defines $\hat{p}_\alpha$, and $\zeta_\alpha(t)$ is a Gaussian white noise with zero mean and variance $2D_r$.
%Thus Eq.~(\ref{eq:4}) and Eq.~(\ref{eq:5}) are the governing equations of motion for each bead. 
We nondimensionalized the model by choosing the units of length and time as $l_0$ and ${c}/{K_s}$, respectively. Thus, four dimensionless parameters mainly determine the collective dynamics: non-dimensional intracellular pressure ($\Tilde{P}=P/K_s$), adhesion strength ($\Tilde{K}_{adh}=K_{adh}/K_s$), self-propulsion speed ($\Tilde{v}_0={cv_0}/{K_sl_0}$), and rotational noise strength ($\Tilde{D}_r={D_rc}/{K_s}$). We used the model to simulate a nearly confluent tissue with 256 cells, each consisting of 50 beads. Starting with random cell centers, the tissue monolayer was simulated using Euler's method up to $10^7$ iterations with a time step $10^{-3}$, under periodic boundary conditions (see SI for simulation details and Table S1 for parameter values).

\section{\label{sec:level3} RESULTS}

We first explored how the cell shapes and tissue state qualitatively change with the variations of cell-cell adhesion ($\Tilde{K}_{adh}$) and rotational noise strength ($\Tilde{D}_r$) (see Fig.~\ref{fig1}B). When $\Tilde{K}_{adh}$ and $\Tilde{D}_r$ are low, cells are mostly elliptical and rounded shaped with noticeable intercellular gaps (Fig.~\ref{fig1}B, bottom-left). With an increase in $\Tilde{K}_{adh}$ or $\Tilde{D}_r$, regular polygonal cell shapes (like hexagons and pentagons) coexist with rounded shapes, along with some intercellular gaps in the tissue (Fig.~\ref{fig1}B, top-left and bottom-right). In contrast, when both $\Tilde{K}_{adh}$ and $\Tilde{D}_r$ are high, almost all cells become regular hexagonal-shaped with virtually no cell-cell gaps (Fig.~\ref{fig1}B, top-right), suggesting cell jamming. 

\begin{figure}[hbt!]
\includegraphics[width=8.6cm]{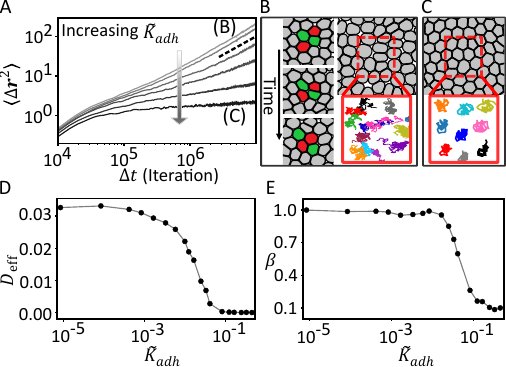}% Here is how to import EPS art
\caption{\label{fig:epsart} \textbf{Fluid to solid transition with increasing intercellular adhesion.} 
\textbf{(A)} MSD of cell centers for different values of $\tilde{K}_{adh}$ (top to bottom: $\tilde{K}_{adh} = 8.3 \times10^{-5},8.3 \times10^{-3}, 0.025, 0.042, 0.083, 0.25$), showing diffusive to subdiffusive and caged behavior as $\tilde{K}_{adh}$ increases. The dashed line indicates a slope of 1 on the log-log plot. \textbf{(B-C)}
Zoomed-in snapshots of cell collectives and the cell center trajectories at low (B) and high (C) adhesion strengths (corresponding to extreme values of $\tilde{K}_{adh}$, denoted by (B) and (C) in Panel A). Regular
polygonal shapes arise in
C, but mostly rounded shapes
emerge in B with noticeable intercellular gaps. Cell center
trajectories look diffusive in B, while they appear caged in C.
 In B (left), red and green cells show spontaneous neighbor exchange (T1 transition) in the liquid phase. \textbf{(D)} The effective diffusivity, quantified as an order parameter, is shown against
$\tilde{K}_{adh}$. \textbf{(E)} MSD exponent, as a function of $\tilde{K}_{adh}$. Parameters: $\tilde{v}_0 = 16.6\times10^{-3}$,$\tilde{P} = 0.2$, $\tilde{D}_r = 5.2\times10^{-4}$. Other parameters are from Table S1. }
\label{fig2}
\end{figure}

To characterize the above phenomenon, we first evaluated the mean-squared displacement (MSD) of cell center trajectories, defined as
$\langle{\Delta\textbf{\textit{r}}^2(\Delta{t})}\rangle =  \langle [\textbf{\textit{r}}_\alpha^{cm}(t_0+\Delta{t}) - \textbf{\textit{r}}_\alpha^{cm}(t_0) ]^2 \rangle$. Here, $\textbf{\textit{r}}_\alpha^{cm}(t)$ is the position vector of the $\alpha$-th cell center at time $t$, $t_0$ is a reference time, $\Delta{t}$ is the lag time, and the angular brackets denote the average over all cells and over different ensembles. We found that the cell motion is diffusive for low $\Tilde{K}_{adh}$ (see Fig.~\ref{fig2}A). However, with the increasing $\Tilde{K}_{adh}$, the motion becomes subdiffusive, and eventually, the MSD plateaus at large times, indicating the arrested motion of cells caged by their neighbors. Thus, the monolayer transforms from a fluid to a solid regime with increasing adhesion (when other parameters are fixed). The monolayer also gets solidified with increasing $\Tilde{P}$ or $\Tilde{D}_r$, and decreasing $\Tilde{v}_0$ (Fig. S3).

%%%%%%%%%%%%%%%%%%%%%%%%%%%%%%%%%%%%%%%%%
%%%%%%%%%%%%%%%%%%%%%%%%%%%%%%%%%%%%%%%%%%%%%%%%%%%%%

We further observed that cells spontaneously exchange their neighbors (known as T1 transition \cite{duclut2022active}) in the fluid phase (see Movie S1 and Fig.~\ref{fig2}B). Correspondingly, the cell center trajectories appear diffusive in nature (Fig.~\ref{fig2}B). In contrast, the trajectories look caged in the solid phase, as expected from the MSD curves (Fig.~\ref{fig2}A). We then computed an effective diffusivity from the long-time MSD behavior \cite{Bi2016}, defined by
 $D_{\text{eff}} =  \lim_{t\to\infty} \langle{\Delta\textbf{\textit{r}}^2(\Delta{t})}\rangle / {4D_0\Delta{t}}$, where $D_0 = {v_0}^2/2D_r$ can be regarded as the self-diffusivity for an isolated single cell \cite{Bi2016}. The order parameter, $D_\text{eff}$, showed a transition with increasing intercellular adhesion (Fig.~\ref{fig2}D). We also measured the MSD exponents ($\beta$) by fitting the MSD curves at large times with the function: $\langle{\Delta\textbf{\textit{r}}^2(\Delta{t})}\rangle \sim \Delta{t}^\beta$. The  MSD exponent was $1$ for the fluid regime and showed a sharp drop as the tissue solidified (Fig.~\ref{fig2}E).  

%%%%%%%%%%%%%%%%%%%%%%%%%%%%%%%%%%%%%%%%%%%%%%%%%
\begin{figure}[hbt!]
\includegraphics[width=8.6cm]{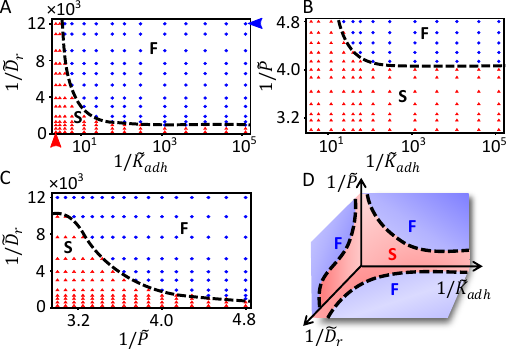}% Here is how to import EPS art
\caption{\label{fig:epsart} \textbf{Tissue phase diagrams.} \textbf{(A-C)} Phase diagrams are shown in the 2D parameter space of   $1/\tilde{D}_r$ vs. $1/\tilde{K}_{adh}$ (A), $1/\tilde{P}$ vs. $1/\tilde{K}_{adh}$ (B), and $1/\tilde{D}_r$ vs. $1/\tilde{P}$ (C). Blue dots represent the fluid phase with $D_\text{eff} > 0.001$, and red triangles represent the solid phase with $D_\text{eff} \leq 0.001$. Arrowheads in panel A correspond to extreme opposite phases that are discussed in Fig. 4. \textbf{(D)} The 2D phase planes are organized into a schematic 3D phase diagram. Letters `F' and `S' denote the fluid and solid phases, respectively.
 Parameters: $\tilde{v}_0 = 16.6\times10^{-3}$, $\tilde{P} = 0.2$, $\tilde{D}_r = 5.2\times10^{-4}$, $\tilde{K}_{adh} = 8.3\times10^{-4}$ (Note that two of these parameters are varied in panels A-C). Other parameters are from Table S1.}
\label{fig3}
\end{figure}
%%%%%%%%%%%%%%%%%%%%%%%%%%%%%%%%%%%%%%%%%%%%%%%

Next, we explored the phase space spanned by the three dimensionless parameters, the intercellular adhesion  ($\Tilde{K}_{adh}$),  the intracellular pressure ($\Tilde{P}$),  and the rotational noise strength ($\Tilde{D}_r$).  We determined the phase diagrams in the 2D parameter space by varying two of the above parameters (Fig.~\ref{fig3}A-C). We identified the fluid phase 
by the cut-off $D_\text{eff} > 0.001$ and solid by $D_\text{eff} \leq 0.001$ (cut-off values were chosen following Fig.~\ref{fig2}D). 
Here, we determined the phase diagrams by keeping the cell motility ($\tilde{v}_0$) fixed, though the phase planes can also be constructed by varying $\tilde{v}_0$ (Fig.\;S4).
The 2D phase planes can be combined to visualize a schematic 3D phase diagram (Fig.~\ref{fig3}D). Similar to the SPV model prediction \cite{Bi2016}, Fig.~\ref{fig3}D shows that the monolayer gets fludized with  
increasing persistence time scale for the rotational noise (given by $1/\Tilde{D}_r$). But, concerning the effect of cell-cell adhesion ($\Tilde{K}_{adh}$),
our phase diagram fundamentally differs from Vertex and SPV models \cite{Bi2016, park2016collective},  where adhesion helps fluidization in a confluent tissue. However, our prediction agrees with the speculated jamming phase diagrams derived from previous experiments \cite{Sadati2013, ilina2020cell}, which hypothesized that intercellular adhesion solidifies the tissue. In addition, our phase diagram crucially points out that solidification can occur with increasing intracellular pressure, consistent with the observation that high cytoplasmic pressure promotes epithelial integrity \cite{jones2021cytoplasmic}.

%%%%%%%%%%%%%%%%%%%%%%%%%%%%%%%%%%%%%%%%%%%%%%%%%%%%%%%%%%%%%%%%%%%%%%%%
\begin{figure*}[hbt!]
\includegraphics[width=18.2cm]{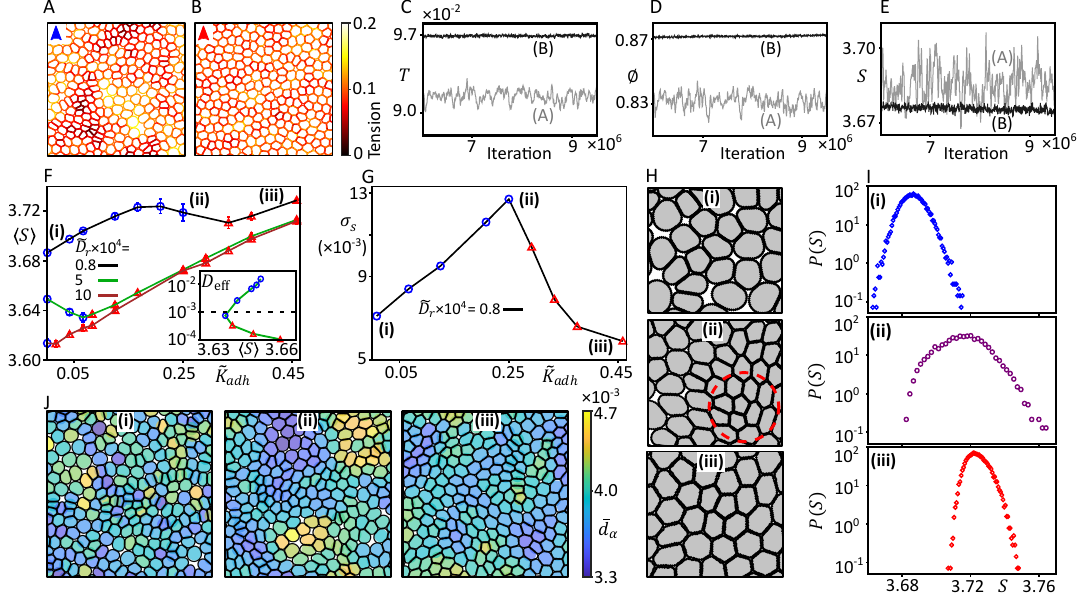}% Here is how to import EPS art
\caption{\label{fig:epsart} \textbf{Tension and shape fluctuations in tissue monolayers.} \textbf{(A-B)} Heat maps showing cellular edge tensions in the tissue kept in fluid (A) and solid (B) phases, as shown by arrowheads in the phase diagram of Fig. 3A. Parameters: $\tilde{K}_{adh} = 8.3\times10^{-6}$, $\tilde{D}_r = 0.8\times10^{-4}$ for A and $\tilde{K}_{adh} = 0.42$, $\tilde{D}_r = 0.001$ for B. Other parameter values are as mentioned in Fig. 3 caption.  \textbf{(C-E)} For a single tissue, the average tension (C), area fraction (D), and shape index (E) are plotted over time in the steady state. The grey and black lines denote two opposite phases (fluid and solid, respectively) corresponding to panels A and B. \textbf{(F)}  The shape index, averaged over time and many ensembles, is shown with $\tilde{K}_{adh}$ for various $\tilde{D}_r$ values. Inset: The dynamical order parameter, $D_\text{eff}$, versus $\langle{S}\rangle$ for $\tilde{D}_r\times 10^{4}= 5$. \textbf{(G)} Standard deviation of the shape index showed a peak when plotted with $\tilde{K}_{adh}$. Blue circles and red triangles represent fluid and solid phases, respectively, in F and G. \textbf{(H-I)} Instantaneous zoomed configurations (H) and probability distributions of shape indices (I), corresponding to the marked points, (i), (ii), and (iii) in the panels F, G. In H(ii), regular polygonal cell shapes (within the dashed red circle) coexist with elliptical shapes, suggesting shape fluctuations. In I(ii), the shape index distribution is much broader than the other regions (I(i) \& I(ii)). \textbf{(J)} Each cell is color mapped with the amount of distance it traversed within a time window (from $2\times10^5$ to $2\times10^6$ iterations) in the three specified regions, (i), (ii), and (iii) of Panel G. In J(ii),  cells with large and small displacements coexist together, forming connected clusters. Parameters: $\tilde{K}_{adh} = 8.3\times10^{-4}$ for region-i, $\tilde{K}_{adh} = 0.25$ for region-ii, and $\tilde{K}_{adh} = 0.46$ for region-iii. Other parameters are from Fig.~\ref{fig3} and Table S1.}
\label{fig4}
\end{figure*}
%%%%%%%%%%%%%%%%%%%%%%%%%%%%%%%%%%%%%%%%%%%%%%%%%%%%%%%%%%%%%%%%%%%%%%%%%%%%%%%%

Since cells can modulate cortical tension during morphogenesis \cite{heisenberg2013forces}, we calculated the tension in cell-edges, defined as $T_i = |l_i - l_0|$ (ignoring the multiplicative constant), where $l_i$ is the distance between $i$-th and $(i-1)$-th beads in a cell and $l_0$ is the natural spring length. Deep in the fluid phase, the tension is highly heterogeneous over the tissue (Fig.~\ref{fig4}A), while it is homogeneous in the solid phase (Fig.~\ref{fig4}B). We also calculated the average tension in the steady state (given by  $T = (\sum_{M, N}^{}T_i)/M N$, where $M$ and $N$ are the bead and cell numbers, respectively), and found that it is lower and display higher fluctuations in the fluid phase than the solid phase (Fig.~\ref{fig4}C, Fig. S5). Thus, tension fluctuations decrease with the solidification, as recently shown in the `Dynamic Vertex Model' (DVM) with active tension dynamics \cite{Kim2021}. 

However, tension fluctuation is an emerging property of our model, while in DMV, tension fluctuation was introduced via a dynamical equation for the edge tension.

We also measured the area fraction ($\phi$) over time, defined as $\phi = ({\sum_{\alpha=1}^{N}A_\alpha})/L^2$, where $A_\alpha$ is the area of the $\alpha$-th cell and $L$ is the simulation box length.   The area fraction substantially fluctuates in the fluid phase compared to the solid phase (Fig.~\ref{fig4}D), although it increases marginally (from $0.83$ to $0.87$) from the fluid to solid phase (Fig.~\ref{fig4}D). Thus, the tissue remains near confluent throughout the transition, similar to the measured {\it in vivo} volume fraction (around $80 \% - 90 \%$) in dense zebrafish tissues \cite{Mongera2018}.  
We next computed the cell shape index, ${S} = (\sum_{\alpha=1}^{N}S_\alpha)/N = (\sum_{\alpha=1}^{N}{P_\alpha}/{\sqrt{A_\alpha}})/N$, $P_\alpha$ and $A_\alpha$ being the perimeter and area of the $\alpha$-th cell, respectively \cite{Bi2015, Bi2016}. Note that this shape index can be measured as an emerging property \cite{Bi2015, Bi2016, Kim2021, Chiang2016, Park2015}, while the `target shape index' is a tunable parameter defined in the energy function of Vertex and Voronoi (SPV) models \cite{Bi2015, Bi2016}. We observed that the shape index fluctuates wildly over time in the fluid phase, but its fluctuations diminish in the solid phase (Fig.~\ref{fig4}E).

%%%%%%%%%%%%%%%%%%%%%%%%%%%%%%%%%%%
\begin{figure*}[hbt!]
\includegraphics[width=17.8cm] {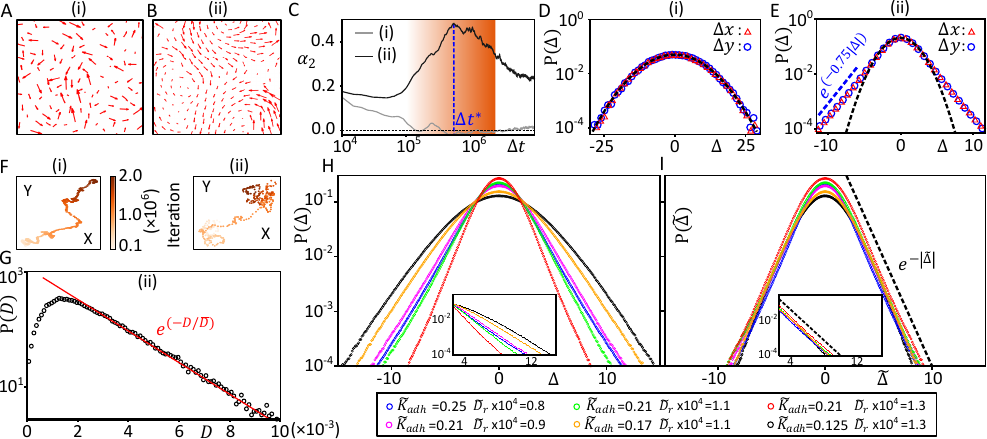}% Here is how to import EPS art
\caption{\label{fig:epsart} \textbf{Dynamic heterogeneity and glassy dynamics.}  \textbf{(A-B)} Cell center displacements over a time window (from $10^5$ to $2\times10^6$ iterations) for low and high adhesion strengths (marked by \textbf{(i)} and \textbf{(ii)}, respectively,  corresponding to  Fig. 4G). For lower adhesion (region-i, A), instantaneous displacements are random and uncorrelated, whereas the displacement field shows swirling patterns for higher adhesion (region-ii, B). \textbf{(C)} The non-Gaussian parameter, $\alpha_2(\Delta{t})$, shows a peak for higher adhesion (region-ii) around the lag time $\Delta{t}^*$. The shaded region spanning $\Delta{t}^*$ indicates the time window where the displacement fields (panels A, B) and trajectories (panel F) were observed. \textbf{(D-E)} Probability distributions of cell center displacements at the lag time $\Delta{t}^*$.  Black dashed lines indicate the best fit Gaussian. The Blue dashed line (in E) shows an exponential fit. \textbf{(F)} Sample cell center trajectories within a time window (corresponding to the shaded region in C). The trajectory is diffusive for lower adhesion (region-i), but a cage rearrangement event (hopping trajectory) was seen for higher adhesion (region-ii). \textbf{(G)} Probability distribution of diffusion coefficients ($D$) measured from time-averaged MSD curves of individual cells for the region-ii. The red line is an exponential fit. \textbf{(H-I)} Probability distributions of cell center displacements (H) and scaled displacements (I) for different values of $\tilde{K}_{adh}$ and $\tilde{D}_r$. Note that the tails of all distributions of normalized displacements (defined by $\tilde\Delta = 1/\sqrt{(2 \bar{D} \Delta{t^*})}$) follow a single master curve with exponent 1 (indicated by the dashed line in I). Insets of H and I show zoomed tail parts of corresponding distributions.  The parameters are the same as in Fig.~\ref{fig4}.    
}
\label{fig5}
\end{figure*}
%%%%%%%%%%%%%%%%%%%%%%%%%%%%%%%%%%%%%%%%%

The Vertex and SPV models \cite{Bi2015, Bi2016} showed that the average shape index, $\langle{S}\rangle$, serves as a structural order parameter. In DVM and SPV models \cite{Bi2016, Kim2021}, $\langle{S}\rangle$ increases monotonically as the tissue fluidizes. In contrast, our model reveals that $\langle{S}\rangle$ behaves non-monotonically with adhesion near the fluid-solid transition, especially for lower $\tilde{D}_r$ (Fig.~\ref{fig4}F). Moreover, the dynamical order parameter, $D_\text{eff}$, is also non-monotonic with $\langle{S}\rangle$  (Fig.~\ref{fig4}F inset), indicating that both fluid and solid phases can possess the same $\langle{S}\rangle$ value close to the transition. Thus, the shape index alone cannot distinguish between the tissue phases. 
We then examined how the shape index fluctuates across the transition, as marked by \textbf{(i)}, \textbf{(ii)}, and \textbf{(iii)} in Fig.~\ref{fig4}F (denoting three distinct regions: deep in the fluid phase, near the phase boundary, and the solid phase, respectively). The standard deviation of $S$ ($\sigma_S = \sqrt{\langle{S^2}\rangle - \langle{S}\rangle^2}$) displayed a peak near the transition boundary in region-ii (Fig.~\ref{fig4}G). Corresponding tissue configurations are shown in Fig.~\ref{fig4}H, and the distributions of the shape index ($P(S)$) are plotted in Fig.~\ref{fig4}I. At region-ii, where $\sigma_S$ has a peak, the distribution $P(S)$ becomes more skewed and broader than other regimes (see Fig.~\ref{fig4}I). Interestingly,  elongated polygonal and rounded cell shapes coexist in region-ii (Fig.~\ref{fig4}H\textbf{(ii)}), whereas cell shapes are mostly rounded in region-i and polygonal (i.e., pentagonal or hexagonal) in region-iii (Fig.~\ref{fig4}H). Moreover, in the solid regime (region-iii), the monolayer resembles a tightly packed foam, where the normalized cell area follows a generalized k-gamma distribution as found earlier in granular material \cite{aste2008emergence}, foam \cite{katgert2010jamming}, and the Vertex model \cite{sadhukhan2022origin} (see Fig. S6).

The cell shape heterogeneity that arises near the transition (Fig.~\ref{fig4}G, H) may be linked with underlying dynamic heterogeneity in cell motion.
To examine this, within a time window ($t_1$ to $t_2$), we measured the total normalized distance covered by individual cell centers, $\bar{d_\alpha}$, defined for the $\alpha$-th cell as $\bar{d_\alpha} = d_\alpha/(\sum_{\alpha=1}^M{d_\alpha})$, where $d_\alpha = \sum_{t_i=t_1}^{t_2-\Delta{t}}\Delta{r}_\alpha(t_i) = \sum_{t_i=t_1}^{t_2-\Delta{t}}|\textbf{\textit{r}}_\alpha^{cm}(t_i+\Delta{t}) - \textbf{\textit{r}}_\alpha^{cm}(t_i)|$. As suggested by the heatmaps of Fig.~\ref{fig4}J, different mobilities exist in region-i but in a spatially random fashion (Fig.~\ref{fig4}J(\textbf{i})). However, cells in region-iii mostly have lower mobilities (Fig.~\ref{fig4}J(\textbf{iii})). Significantly, in region-ii, fast-moving and slow-moving cells form clusters (Fig.~\ref{fig4}J(\textbf{ii})), suggesting that cells with distinct mobilities coexist with spatial correlation. This indicates that cells are dynamically heterogeneous, suggesting a glass transition, as found in colloidal glasses \cite{weeks2002properties, tah2021understanding} and in dense bacterial biofilm \cite{bacterialglass2024}.

 To probe the glassy versus fluidic behavior, we plotted the vector field of cell center displacements for region-i and region-ii, corresponding to lower and higher adhesion strengths, respectively (see Fig.~\ref{fig5}A, B). The displacement vectors are random for region-i but exhibit swirling patterns indicating spatial correlations in region-ii. Also, large and small displacements coexist in the vector field of region-ii, suggesting dynamic heterogeneity similar to polycrystalline materials and glasses \cite{biswas2013micromechanics, weeks2000three, weeks2002properties, nagamanasa2011confined}. Notably, similar swirling displacement fields were observed in previous models \cite{bameta2012broad, Bi2016, Mandal2017} and epithelial monolayers \cite{Garcia2015, Angelini2011, poujade2007collective, petitjean2010velocity}.

Another quantity that can capture the glassy behavior is the 
non-Gaussian parameter defined by 
$\alpha_2(\Delta{t}) = \langle{\Delta\textbf{\textit{r}}^4(\Delta{t})}\rangle/(2 {{\langle{\Delta\textbf{\textit{r}}^2(\Delta{t})}\rangle}^2}) - 1$. It is non-zero if displacement distributions deviate from Gaussian \cite{Chiang2016, weeks2000three}. In Fig.~\ref{fig5}C, $\alpha_2$ is almost zero in region-i (lower adhesion), for which diffusive cell dynamics is expected. In contrast, $\alpha_2$ has a sharp non-zero peak at the lag time $\Delta{t^*}$ in region-ii (higher adhesion), for which cell dynamics is subdiffusive. Accordingly, the displacement distribution $P(\Delta)$, measured at $\Delta{t^*}$, is Gaussian in region-i, but it develops broad exponential tails in region-ii (Fig.~\ref{fig5}D, \ref{fig5}E). Such non-zero peaks in $\alpha_2$ and corresponding non-Gaussian tails in displacement distributions have been observed in inanimate glassy materials  \cite{weeks2000three, weeks2002properties, nagamanasa2011confined}, and also in living systems like embryonic tissues \cite{schoetz2013glassy} and cytoskeletal dynamics in human muscle cells \cite{bursac2005cytoskeletal}.

The broad exponential tail in the displacement distribution (Fig.~\ref{fig5}E) represents the motile cells performing random jumps between cages \cite{weeks2000three, weeks2002properties, bursac2005cytoskeletal}. The lag time $\Delta{t^*}$, at which $\alpha_2$ shows a  peak (Fig.~\ref{fig5}C), gives the cage breaking time scale \cite{weeks2000three, weeks2002properties, nagamanasa2011confined}. Indeed, within a time window spanning $\Delta{t^*}$, a sample trajectory in region-ii displays rare and quick hopping of a cell between cages formed by its neighbors (Fig.~\ref{fig5}F(ii)), but the trajectory looks simply diffusive in region-i (Fig.~\ref{fig5}F(i)).

 However, what could be the physical origin of the exponential tails in displacement distributions observed in the glassy regime (region-ii)? We found that the individual diffusion coefficients ($D$),  measured from the time-averaged MSD curves, have a broad skewed distribution that has an exponential-like form (Fig.~\ref{fig5}G). Thus, the exponential tails in displacement distributions (Fig.~\ref{fig5}H) can stem from the cell-to-cell heterogeneity in individual diffusivities \cite{he2016dynamic, wang2012brownian}. Suppose individual cells in the ensemble follow independent diffusive dynamics characterized by Gaussian displacement distributions but with an exponentially distributed diffusion coefficient, $D$. Then, non-gaussian displacements can be achieved as the convolution of Gaussian --- We can write the displacement distribution (at a lag time $\Delta{t^*}$) as: 
$P(\Delta{x}) = \int {p(D)}\, g(\Delta{x}|D)\, dD$, 
where, $g(\Delta{x}|D) = (1/\sqrt{4\pi{D}\Delta{t^*}})\exp(-\Delta{x^2}/4D\Delta{t^*})$.  
If we assume the distribution of $D$ as $p(D) = (1/\bar{D})\exp(-D/\bar{D})$, where $\bar{D} = (\sum_{\alpha=1}^N{D})/N$ is the mean value of $D$ averaged over all cells, we obtain the form: $P(\Delta{x}) = (1/2\sqrt{\bar{D} \Delta{t^*}})\exp(\Delta{x}/\sqrt{ \bar{D} \Delta{t^*}}) \sim \exp(-\Delta{x}/\sqrt{ \bar{D} \Delta{t^*}})$. Further, note that we numerically obtained individual diffusion coefficients by the definition $D = \lim_{t\to\infty} \langle{\Delta\textbf{\textit{r}}^2(\Delta{t})}\rangle / {4\Delta{t}}$ for the 2-dimensional dynamics. Thus, for the 1D component of displacements ($\Delta x$ or $\Delta y$), we take the mean diffusivity to be $2 \bar D$ to consider the dimensionality. Therefore, if we consider the scaled 1D displacement $\tilde\Delta = \Delta/\sqrt{2 \bar{D} \Delta{t^*}}$, the distribution $P(\tilde \Delta)$ then follows an exponential distribution with exponent $1$. 
 Indeed, for various values of $\tilde{K}_{adh}$ and $\tilde{D}_r$, different non-Gaussian exponential tails with distinct exponents collapse onto a single master curve ($\sim exp(-|\tilde \Delta|$) when the scaled displacement ($\tilde\Delta$) is used (see Fig.~\ref{fig5}H, I). Thus, $\sqrt{2 \bar{D} \Delta{t^*}}$ represents a mean diffusion length by which displacements can be normalized to obtain a universal form, elucidating the link between exponential tails and the heterogeneous diffusivities of individual cells.

\section{\label{sec:level4} DISCUSSION}

{\bf Summary and relevance:} Our result of tissue solidification with adhesion, though contrasts the prediction of Vertex and Voronoi models \cite{Bi2015, Bi2016}, are in line with \textit{in vitro} experiments \cite{Garcia2015, choi2022cell, ilina2020cell}. This also explains how the downregulation of adhesion molecules leads to {\it in vivo} tissue fluidization during zebrafish tail development \cite{Das2017, Mongera2018}. Similar to the experiment \cite{Mongera2018}, adhesion-induced solidification accompanies a marginal change in packing fraction in our model. 
We also found tissue fluidization with increasing self-propulsion speed that captures a recent experiment on motility-driven unjamming in gastrulation \cite{pinheiro2022morphogen}. 
We further show that the average shape index alone cannot be used as a structural order parameter to distinguish the phases due to large shape fluctuations near the transition point. Also, increased adhesion results in dynamic heterogeneity and glassy behavior near the phase boundary.

Our force-based framework, which is conceptually straightforward to implement, relies on a free energy function that obeys the Young-Laplace equation at the level of a single cell in equilibrium (see Section III, SI) --- this is distinct from the topological energy functions used in Vertex, Voronoi, and Cellular Potts models \cite{Bi2015, Bi2016, Chiang2016}. Contrasting these models, we do not assume any target shape parameters since there is no experimental support and such parameters are difficult to physically interpret. Our model exhibits spontaneous neighbor exchange (T1 transitions) unlike the Vertex/Voronoi frameworks, where T1 transitions are performed using an ad hoc criterion of a threshold value on cellular edge lengths \cite{Bi2015, Kim2021}. Further, intercellular gaps naturally arise in our model, unlike the Voronoi tesselation framework, where cell-cell gaps are either completely absent \cite{Bi2015, Bi2016} or implemented by assuming extra `non-physical' vertices \cite{Kim2021}.  
We also observe the spontaneous formation of multicellular rosettes (where five or more cells share a vertex) \cite{harding2014roles} (Fig. S7). In contrast,  rosettes were formed in Vertex models by implementing an ad hoc `edge collapse' procedure that randomly reduces edge lengths to zero \cite{yan2019multicellular}.

{\bf Connection with other models: }
To compare with the jamming in passive foams, we studied our model without the self-propulsion and adhesion ($v_0 = 0$, $K_{adh} = 0$). In a rigid box, we produced jammed states by inflating the cells with increasing pressure (i.e., with increasing packing fraction). We show that the tissue becomes confluent when the normalized area fraction reaches unity ($\phi/\phi_{max} = 1$) at an average shape index $\langle{S}\rangle \approx 3.81$ (see Fig. S8 and Section V, SI). This is similar to the crowding-dependent jamming in the Deformable Particle (DP) model \cite{dp2018prl}. Interestingly, the Vertex and SPV models \cite{Bi2015, Bi2016} also show a liquid-to-solid transition at $\langle{S}\rangle \approx 3.81$.

Note that the level of confluency can be described by $\phi/\phi_{max}$ ($\phi/\phi_{max} = 1$ implies full confluency) \cite{dp2018prl}. 
Though the absolute packing fraction cannot reach unity in our particle-based model (unlike Vertex and SPV models), we maintained a near-confluent regime, where $(\phi/\phi_{max}) \sim 0.94-1$ across the transition (corresponding to Fig. 2-4). This is similar to the
{\it in vivo} experiment on zebrafish tissues that show 80\%-90\% confluency \cite{Mongera2018}.

However, to compare with Vertex and SPV models, we performed the Voronoi tessellation of space (using the cell centers) to obtain the Voronoi polygons corresponding to the cells (Fig. S9A) and measured the shape indices of the Voronoi cells, $\langle{S}\rangle_{Vor}$ (see Section VI, SI). We found that the liquid-solid transition takes place at $\langle{S}\rangle_{Vor} \approx 3.81$ for low values of $D_r$, similar to the Vertex and SPV models \cite{Bi2015, Bi2016}, though the transition point deviates from $3.81$ for higher $D_r$ (Fig. S9B-C). Moreover, we achieved very high confluency by increasing the rest length of cell edges (represented by Hookean springs) and found that the packing fraction hardly altered during the liquid-solid transition ($\phi \sim 0.891 - 0.893$, i.e., $(\phi/\phi_{max}) \sim 0.998-1$),  when adhesion was increased from zero (Fig. S10C-E). The average coordination number also stayed very high and almost constant ($Z \sim 5.90 - 5.98$) during the transition (Fig. S10F). This indicates an adhesion-dependent active glass transition in the confluent limit, not a crowding-dependent jamming. Thus, tissue solidification arises from dampening the active motility as higher adhesion leads to increased resistance to shear forces between cells. Further, measuring $\langle{S}\rangle_{Vor}$ from Voronoi cells again showed that the transition happens at $\langle{S}\rangle_{Vor} \approx 3.81$  (Fig. S10G), exhibiting the behavior of Vertex and SPV models \cite{Bi2015, Bi2016}. Nevertheless, the actual shape indices measured from cell contours are non-monotonic across the transition (Fig. 4F, S10H).

%Notably, our active bead-spring model is a force-based framework that is conceptually straightforward and intuitive to implement and captures many realistic features --- this is distinct from the topological energy-based models like Vertex, SPV, and Cellular Potts models \cite{Bi2015, Bi2016, Chiang2016}. First, unlike the restricted shape fluctuations in Vertex and SPV models, we describe deformable cell shapes by many degrees of freedom (multiple beads and springs). Second, we do not assume any target shape parameters since no experimental foundation supports such a consideration, and such parameters are difficult to interpret physically. Third, intercellular gaps spontaneously arise in our model, unlike the models based on Voronoi tesselation where cell-cell gaps are either completely absent \cite{Bi2015, Bi2016} or included on an ad hoc basis \cite{Kim2021}. Fourth, neighbor exchanges occur naturally in our model through spontaneous T1 transitions, unlike the Vertex and Voronoi frameworks, where T1 transitions are performed using an ad hoc criterion of a threshold value on cellular edge lengths \cite{Bi2015, Bi2016, Kim2021}. Another observed feature in organ development and morphogenesis is the formation of multicellular rosettes (where five or more cells share a vertex) \cite{harding2014roles}. 

As shown before, in the passive limit, our active model is similar to passive deformable particle (DP) models \cite{Mkrtchyan2014,dp2018prl}. However, including `active and noisy' motility is crucial to observe the glass transition. Though a few studies included tunable confluency in the Voronoi framework \cite{Kim2021, levine2018pre}, our implementation is distinct and straightforward, inspired by DP models. There are also a few active models of deformable cells like ours \cite{ringpolymer_softmatt2022, Jiayi_2022_BPJ}. However, the intracellular pressure in these models depends on the instantaneous cell area, which is similar to vertex-based models but distinct from our constant pressure framework that aligns with {\it in vivo} measurement of osmotic pressure in tissues \cite{otgarcampas_osmoticpressure2023}. Recently, a computational package \cite{polyhoop2024} has been developed to simulate various soft materials, including biological tissues, foams, bubbles, etc. However, this framework is overtly complex (with 23 parameters), while ours is a simple coarse-grained framework with a few parameters corresponding to experiments (such as adhesion, intracellular pressure, edge tension, etc.).

{\bf Application and future directions:} 
Intriguingly, despite having a distinct single-cell energy function compared to the DP, Vertex, and SPV models, our model still shows the liquid-solid transitions at $\langle{S}\rangle_{Vor} \approx 3.81$ in certain limits. This may be tied to the argument of isostaticity based on the polygonal tiling of the 2D space, leading to an isostatic coordination number of $Z_{iso}=5$ (corresponding to the $\langle{S}\rangle_{Vor} = 3.81$ for a pentagon) as explained in \cite{Bi2015}. However, a detailed correspondence between the vertex-based framework and ours would be an interesting future problem. Since real biological tissues are near-confluent, our model is suitable for studying the effect of confluency on active glassy behaviors. In near-confluent systems, it is worthwhile to study the interplay of two mechanisms of solidification: crowding-dependent jamming (in sub-confluent regimes) and adhesion-driven or tension-dependent rigidity transition (in full confluent regimes) \cite{lawson2021jamming}. For instance, a recent study on Xenopus development \cite{weng2022convergent} shows the signatures of both mechanisms. Another work has 
 discovered that periodic assemblies of myosin motors interconnect with the actin cortex and act like a chain of sarcomeric units in each epithelial cell \cite{seham2013currbio}. Such an arrangement resembles our bead-spring loops representing the cell cortex. The authors found that the distance between sarcomeric units changes upon perturbation of the actin cortex, and it relaxes to the original state when the perturbation is removed. Our model could be suitable to understand the effect of such mechanical perturbation on cell shape and tissue rigidity. Moreover, a detailed continuum model has recently explored how T1 transitions emerge from stochastic formation and dissociation of cell-cell adhesive bonds at the level of a few cells \cite{apposed_ploscompbio2022}. Since T1 transitions spontaneously emerge at the tissue level in our coarse-grained model, an interesting question would be how the T1 transitions depend on confluency and adhesion.

%\section*{DISCUSSION}
\begin{acknowledgments}
 DD and SR thank DBT (Government of India, Project No. BT/RLF/Re-entry/51/2018) and IISER-Kolkata for financial support. We also thank Dapeng Bi (Northeastern University), Dibyendu Das (IIT Bombay), Anirban Sain (IIT Bombay) and Somajit Dey (IISER Kolkata) for their useful discussions. 
\end{acknowledgments}

%\bibliography{ref.bib}

%%%%%%%%%%% SI Text %%%%%%%%%%%%%%%%%%%%%%%%%%
\pagebreak
\onecolumngrid

\newpage

\begin{center}
\textbf{\large Supplementary Information: \\Role of intercellular adhesion in modulating tissue fluidity}
\end{center}

\setcounter{equation}{0}
\setcounter{figure}{0}
\setcounter{table}{0}
\setcounter{section}{0}
\setcounter{page}{1}
\makeatletter
%%%%%%%%%%%%%%%%%%%%%%%%%%%%%%%%%%%%%%%%%%%%%%

\renewcommand{\figurename}{FIG.} % for figure name
\renewcommand{\thefigure}{S\arabic{figure}} % for figure numbering format (e.g. here S1, S2 etc.)

\section{\label{sec:level1} MODEL DETAILS} % \lowercase{via} \textbackslash\textbackslash}

\subsection{\label{sec:level1A}{Force inherent to an isolated cell}}
\begin{figure}[hbt!]
\includegraphics[width=8.6cm]{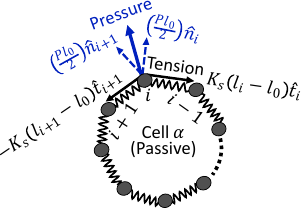}
\caption{ \textbf{Bead-spring model of a single cell.} The force components due to the tangential spring tension are shown in black arrows. The blue arrows denote the normal components of the pressure force. The neighbors of the $i$-th bead are the $(i-1)$-th and $(i+1)$-th beads.}
\label{fig:S1}
\end{figure}

Motivated from previous studies \cite{Astrom2006pre, Mkrtchyan2014, Madhikar2021}, a single cell is modeled as a closed loop of beads connected through elastic springs with stiffness constant $K_s$ and natural spring length $l_0$ (Fig.~\ref{fig:S1}). In our simulation, each cell is made up of 50 beads. Each bead experiences tangential tension forces from the two adjacent neighbor springs, and an internal pressure force, $P l_0$, directed (outward) normal to the line tension of the given spring. 
(Fig.~\ref{fig:S1}). \\

In a coarse-grained view, these intracellular forces describe the roles of the actomyosin cortex and cytoplasmic pressure that give the cell its structural integrity. The intracellular force for the $i$-th bead can be explicitly written as follows:
\begin{align}
\boldsymbol{F}_i^{intra} =\;\; {K_s}({l_i}-{l_0})\hat{t}_i - {K_s}({l_{i+1}}-{l_0})\hat{t}_{i+1}
+ \frac{P{l_0}}{2}(\hat{n}_i + \hat{n}_{i+1})
\label{eq:intraforce}
\end{align}
Here, $l_i$ is the bond length between $i$-th and $(i-1)$-th bead pair, and $l_{i+1}$ is the bond length between $(i+1)$-th and $i$-th beads, respectively. Also, $\hat{t}_i$ and $\hat{t}_{i+1}$ are the tangential unit vectors, and $\hat{n}_i$ and $\hat{n}_{i+1}$ are the corresponding outward normal unit vectors perpendicular to $\hat{t}_i$ and $\hat{t}_{i+1}$, respectively (see Fig.~\ref{fig:S1}). The first two terms represent the total spring-tension force, and the third term represents the outward pressure force on the $i$-th bead.

\subsection{\label{sec:level1B}{Effect of cell-cell interactions}}
 The intercellular force arising due to cell-cell interactions consists of two parts: 
 (i) the spring-like attractive force, which resembles the cell-cell adhesion through adhesive proteins (like E-cadherins), and (ii) the spring-like repulsion that prevents the interpenetration among neighboring cells. Two beads of two different cells would feel the force of adhesion or repulsion only when the Euclidean distance between them is below the cut-off range of the attractive or repulsive forces. Thus the interaction force on the $i$-th bead of $\alpha$-th cell due to the $j$-th bead of $\beta$-th cell is as follows:

%\resizebox{0.95
%\linewidth}{!}{
%\begin{minipage}{\linewidth}
\begin{align}
\boldsymbol{F}_{ij,\alpha\beta}^{inter} &= K_{adh}({r_c}^{adh}-r_{ij,\alpha\beta})\hat{r}_{ij,\alpha\beta}\;\; , \text{if}\;\; {r_c}^{rep}\leq r_{ij,\alpha\beta}\leq {r_c}^{adh} \nonumber\\
&=-K_{rep}({r_c}^{rep}-r_{ij,\alpha\beta})\hat{r}_{ij,\alpha\beta}\;\; , \text{if} \;\;r_{ij,\alpha\beta}< {r_c}^{rep} \nonumber\\
&= 0 \;\;\; ,\;\;\; \text{otherwise}
\label{eq:interforce1}
\end{align}
%\end{minipage}}

%\begin{align}
%\Vec{F}^{ij,\alpha\beta}_{inter} &= K_{adh}({r_c}^{adh}-r_{ij}^{\alpha\beta})\hat{r}_{ij}^{\alpha\beta}\;\; , if\;\; {r_c}^{rep}\leq r_{ij}^{\alpha\betathe }\leq {r_c}^{adh} \nonumber\\
%&= -K_{rep}({r_c}^{rep}-r_{ij}^{\alpha\beta})\hat{r}_{ij}^{\alpha\beta}\;\; , if \;\;r_{ij}^{\alpha\beta}< {r_c}^{rep} \nonumber\\
%&= 0 \;\;\; ,\;\;\; \text{otherwise}
%\label{eq:2}
%\end{align}

Here, $\boldsymbol{r}_{ij,\alpha\beta} = \frac{\boldsymbol{r}_{i,\alpha} - \boldsymbol{r}_{j,\beta}}{|{\boldsymbol{r}_{i,\alpha}-\boldsymbol{r}_{j,\beta}}|}$ is the unit vector of the Euclidean distance between the two interacting beads. $K_{adh}$ and ${r_c}^{adh}$ are the strength and cut-off range, respectively, for the attractive force, and $K_{rep}$ and ${r_c}^{rep}$ are that of the repulsive force.\\
Now, considering all the contributions from other cells within the interaction ranges, the total force of interaction on the $i$-th bead of the $\alpha$-th cell would be :
\begin{equation}
\boldsymbol{F}_{i,\alpha}^{inter} = \sum_{j\neq{i},\beta\neq{\alpha}} \boldsymbol{F}_{ij,\alpha\beta}^{inter}  
\label{eq:interforce2}
\end{equation}

\subsection{\label{sec:level1C} {Active force term via self-propulsion and Equations of motion}}

In addition to the intracellular and intercellular forces, cells can also move via self-propulsion. As in the self-propelled particle (SPP) model \cite{Henkes2011} and self-propelled Voronoi (SPV) model \cite{Bi2016}, we also incorporate a polarity vector $\hat{p}_\alpha = (\cos\theta_\alpha, \sin\theta_\alpha)$ for every cell. In the direction of $\hat{p}_\alpha$, each bead of a cell exerts a self-generated motility force of magnitude $cv_0$, where $c$ is the coefficient of viscous drag, and $v_0$ is the self-propulsion speed. Thus, in each cell, all the beads have the same polarity direction, but the polarity direction differs from cell to cell. All together, the overdamped equation of motion for the $i$-th bead of the $\alpha$-th cell is: 
\begin{equation}
c\dot{\boldsymbol{r}}_{i,\alpha}=\boldsymbol{F}_{i,\alpha}^{intra} + \boldsymbol{F}_{i,\alpha}^{inter} + c{v_0}\hat{p}_\alpha 
\label{eq:eom_position}
\end{equation}

And, the unit polarity vector undergoes a rotational diffusion described as: 
\begin{align}
\frac{\partial{\theta_{\alpha}}}{\partial{t}} &= \zeta_\alpha(t)\;\; ; \quad \langle\zeta_\alpha(t)\zeta_\beta(t')\rangle = 2D_r\delta(t-t')\delta_{\alpha\beta}, 
\label{eq:eom_theta}
\end{align}

where, $\zeta_\alpha(t)$ is a Gaussian white noise with zero mean and variance $2D_r$, which specifies the timescale for reorientation of the polarity vectors as $1/D_r$. \\
Thus Eq.~(\ref{eq:eom_position}) and Eq.~(\ref{eq:eom_theta}) are the governing equations of motion in our model.

\subsection{\label{sec:level1D} Non-dimensional Parameters}

To non-dimensionalize our governing equations, we take the unit of length as $l_0$ and the unit of time as $\tau=\frac{c}{K_s}$ (where $c$ is the coefficient of viscous drag and $K_s$ is the spring constant for a single cell).
Thus, the non-dimensional position vector 
$(\Tilde{\textbf{\textit{r}}}_{i,\alpha})$ would be :
\begin{align}
\Tilde{\textbf{\textit{r}}}_{i,\alpha}&=\textbf{\textit{r}}_{i,\alpha}/{l_0}\nonumber \\ 
\Rightarrow
\textbf{\textit{r}}_{i,\alpha}&=\Tilde{\textbf{\textit{r}}}_{i,\alpha}{l_0}
\label{eq:nondim_1}
 \end{align}

 And, non-dimensional time $(\Tilde{t})$ would be :
 \begin{align}
  \Tilde{t}&=t/{\tau}\nonumber \\
  &=t{K_s}/{c} \nonumber \\ 
  \Rightarrow t&=\Tilde{t}{c}/{K_s}\;
  \label{eq:non_dim_2}
 \end{align}
 Therefore, replacing the position vectors and time with the non-dimensionalized values in Eq.~(\ref{eq:eom_position}), we obtain
 \begin{align}
   c\frac{d(\Tilde{\textbf{\textit{r}}}_{i,\alpha}{l_0})}{d(\Tilde{t}\frac{c}{K_s})} &= \{{K_s}{l_0}[(\frac{l_i}{l_0}-1)\hat{t}_i - (\frac{l_{i+1}}{l_0}-1)\hat{t}_{i+1}] + \frac{P{l_0}}{2}(\hat{n}_i + \hat{n}_{i+1}){\}} \nonumber \\
   &+ {\{}\sum_{j\neq{i},\beta\neq{\alpha}}{}K_{int}(l_0\Tilde{r}_c-l_0\Tilde{r}_{ij,\alpha\beta})\hat{r}_{ij,\alpha\beta}{\}} + c{v_0}\hat{p}_\alpha \nonumber \\
   \Rightarrow  \frac{d\Tilde{\textbf{\textit{r}}}_{i,\alpha}}{d\Tilde{t}} &= [(\frac{l_i}{l_0}-1)\hat{t}_i - (\frac{l_{i+1}}{l_0}-1)\hat{t}_{i+1}] + {\left(\frac{P}{K_s}\right)}\frac{(\hat{n}_i + \hat{n}_{i+1})}{2} \nonumber \\
   &+ \{\sum_{j\neq{i},\beta\neq{\alpha}}{}\left(\frac{K_{int}}{K_s}\right)(\Tilde{r}_c-\Tilde{r}_{ij,\alpha\beta})\hat{r}_{ij,\alpha\beta}\} + \left(\frac{c{v_0}}{K_sl_0}\right)\hat{p}_\alpha \nonumber \\
   \Rightarrow  \frac{d\Tilde{\textbf{\textit{r}}}_{i,\alpha}}{d\Tilde{t}} &= [(\frac{l_i}{l_0}-1)\hat{t}_i - (\frac{l_{i+1}}{l_0}-1)\hat{t}_{i+1}] + \Tilde{P}\frac{(\hat{n}_i + \hat{n}_{i+1})}{2} \nonumber \\
   &+ \{\sum_{j\neq{i},\beta\neq{\alpha}}{}\Tilde{K}_{int}(\Tilde{r}_c-\Tilde{r}_{ij,\alpha\beta})\hat{r}_{ij,\alpha\beta}\} + \Tilde{v}_0\hat{p}_\alpha
   \label{eq:non_dim_3}
 \end{align}   
 
Here, $K_{int}$ denotes the strength of adhesion or repulsion ($K_{adh}$ or $K_{rep}$) in a compact notation. Also, 
$\Tilde{P}=\frac{P}{K_s}$, $\Tilde{K}_{int}=\frac{K_{int}}{K_s}$ and $\Tilde{v_0}=\frac{cv_o}{K_sl_o}$ are dimensionless pressure, interaction strength, and self-propulsion speed, respectively. Note that Eq.~(\ref{eq:non_dim_3}) is a non-dimensional form of Eq.~(\ref{eq:eom_position}). Treating Eq.~(\ref{eq:eom_theta}) in a similar way, we get

 \begin{align}
\frac{\partial{\theta_{\alpha}}}{\partial{(\Tilde{t}\tau})} &= \zeta_\alpha(\Tilde{t})\;\; ; \Tilde{\zeta_{\alpha}}(\Tilde{t})=\tau\zeta_{\alpha}(t/\tau) \nonumber \\ 
\Rightarrow  \frac{\partial{\theta_{\alpha}}}{\partial\Tilde{t}} &= \zeta_\alpha(\Tilde{t}) \label{eq:non_dim_4}
\end{align} 
\centering {and} \\
 \begin{align}      
\langle\Tilde{\zeta}_\alpha(\Tilde{t})\Tilde{\zeta}_\beta(\Tilde{t'})\rangle &= 2(D_r\tau)(\delta(\Tilde{t}-\Tilde{t'}) = 2\Tilde{D}_r\delta(\Tilde{t}-\Tilde{t'}) \label{eq:non_dim_5}
 \end{align}
 
\begin{justify}
Where, $\Tilde{D}_r={D_rc}/{K_s}$ is the non-dimensional rotational noise strength. Thus, the main non-dimensional parameters of our model are $\Tilde{P}$, $\Tilde{K}_{adh}$, $\Tilde{v}_0$ and $\Tilde{D}_r$ (ignoring $\Tilde{K}_{rep}$, since we kept $\Tilde{K}_{rep} >> \Tilde{K}_{adh}$).
\end{justify}

\section{\label{sec:level2} SIMULATION DETAILS} 
\begin{justify}

We consider a system of $N=256$ cells in a square box of size $L = 460 l_0$. Each cell contains $M = 50$ beads. The bead positions are updated by numerically integrating Eq.~(\ref{eq:eom_position}) and Eq.~(\ref{eq:eom_theta}) using the Euler method. The time step is $10^{-3}$ and the total number of iterations is $10^7$. The simulations are carried out using a periodic boundary condition.

\subsection{\label{sec:level2A}Initialization procedure}
 %\textbf{Initialization procedure:}
 \textbf{Generating nearly confluent tissue configuration:}
 To populate the desired number of cells in a given square box, we choose a smaller initial radius for each cell. In our simulation, the initial radius of each cell is taken as $a = 10 l_0$, and the box length is chosen to be $L = 460 l_0$. To avoid cell-cell overlap during the random generation of cells,  the minimum separation between two cell centers is chosen to be $2 a + r_c^{rep}$. 
 In a square box of fixed length, $L$, we generate two random numbers from a uniform distribution as the X and Y coordinates of each cell center. The cell centers are then checked to meet the criterion of lying inside the box and also have no overlap with already generated cells. If there is an overlap, the cell center coordinates are generated repeatedly till the criteria are met. The $50$ beads that make up the cell boundaries are then positioned in an equidistant manner along the perimeter of each cell such that the spring length between successive beads is $2 \pi a/50$. \\
 
 \textbf{Equilibrium dynamics of the tissue configuration:} The initially generated cell-system is then evolved by integrating Eq.~(\ref{eq:eom_position}) with $\Tilde{v}_0 = 0$, i.e., we evolve the passive system without Eq.~(\ref{eq:eom_theta}). During this time window, the cells grow, and by interacting with each other, they approach a nearly confluent equilibrium configuration (with an area fraction above 0.8). This equilibrium configuration then serves as an initial configuration for simulations of the active system. In all our simulations, we equilibrate the randomly generated configurations up to the total iteration of $5\times10^4$. Also note, to avoid any global drift of the system, we integrate Eq.~(\ref{eq:eom_position}) in the center of mass reference frame, i.e., we use the transformed position vectors for each bead, $\boldsymbol{r}^{' i,\alpha} = \boldsymbol{r}^{i,\alpha} -\boldsymbol{R}^{cm}$, where $\boldsymbol{R}^{cm}$ is the position of the center of mass of the entire tissue monolayer.

\subsection{\label{sec:level2B}Dynamics of the active system}
\textbf{Simulation of the active system:} 
After the passive system attained an equilibrium configuration (described above), we reset the time to zero. The final equilibrated configuration of the passive system is then used as the initial configuration for the active dynamics of the tissues. The system is then evolved for a further $10^7$ iteration steps with a non-zero self-propulsion speed ($\Tilde{v}_0 \ne 0$), and the  dynamics is governed by both Eq.~(\ref{eq:eom_position}) and Eq.~(\ref{eq:eom_theta}) simultaneously.\\

 \textbf{Measurement of observable quantities:} 
 We found that after $10^5$ iterations, the active system entered into a steady state, which was determined by observing the time evolution of the dynamical or structural observables like instantaneous area fraction and the shape index (as defined in the next section). The steady-state was confirmed when the observables fluctuated around a well-defined average value over time (as in Fig. 4(C-E), main text). We took this approximate steady-state reaching time as the reference time ($t_0 = 10^5$ iterations) for measuring the dynamical quantities like MSD and the non-Gaussian parameter (as defined in the next section). \\
 
 \textbf{Code details:} The main simulation as well as codes for analysing the data are all written in FORTRAN 90.  We implemented the `cell linked-list' algorithm \cite{AllenTildsley} (which provides a significant speed-up) in the neighbor search process to compute the intercellular forces. The codes are available in the GitHub repository:  \href{https://github.com/PhyBi/Collective-Cell-Dynamics/tree/legacy}{https://github.com/PhyBi/Collective-Cell-Dynamics/tree/legacy}.
\end{justify}

\renewcommand{\thetable}{S\arabic{table}} % for table numbering format (e.g. here S1, S2 etc.).
\begin{table*}[ht!]%The best place to locate the table environment is directly after its first reference in text
\caption{\textbf{List of system parameters.}}
\begin{ruledtabular}
\begin{tabular}{lll} % here l,c,r denotes the corresponding alignment type.
\textrm{Parameters}&
\textrm{Description}&
\textrm{Value}\\
\hline \\[-1.8ex]
${K}_{adh}$ & Intercellular adhesion strength & Varied: $0.001 - 50$\\
$Pl_0$ & Intracellular pressure force & Varied: $2.5 - 4.0$\\
${v}_0$ & Motility & Varied: $0.09 - 0.25$\\
${D}_r$ & Rotational noise strength & Varied: $0.02 - 1.125$\\
${K}_{rep}$ & Intercellular repulsion strength & 1000\\
${r}^{adh}_c$ & Cut-off distance for intercellular adhesion & $2.8 l_0$\\
${r}^{rep}_c$ & Cut-off distance for intercellular repulsion & $1.8 l_0$\\
$l_0$ & Equilibrium spring length & 0.1\\
$K_s$ & Spring stiffness constant & 120\\
$c$ & Coefficient of viscous drag  & 0.5\\
\end{tabular}
\end{ruledtabular}
\label{tab:tableS1}
\end{table*}

%\clearpage
%\newpage

\section{\label{sec:level3}COMPARISON OF OUR MODEL WITH THE TOPOLOGICAL ENERGY FUNCTION OF VERTEX AND VORONOI MODELS}

\begin{figure}[hbt!]
\centering
\includegraphics[width=6cm]{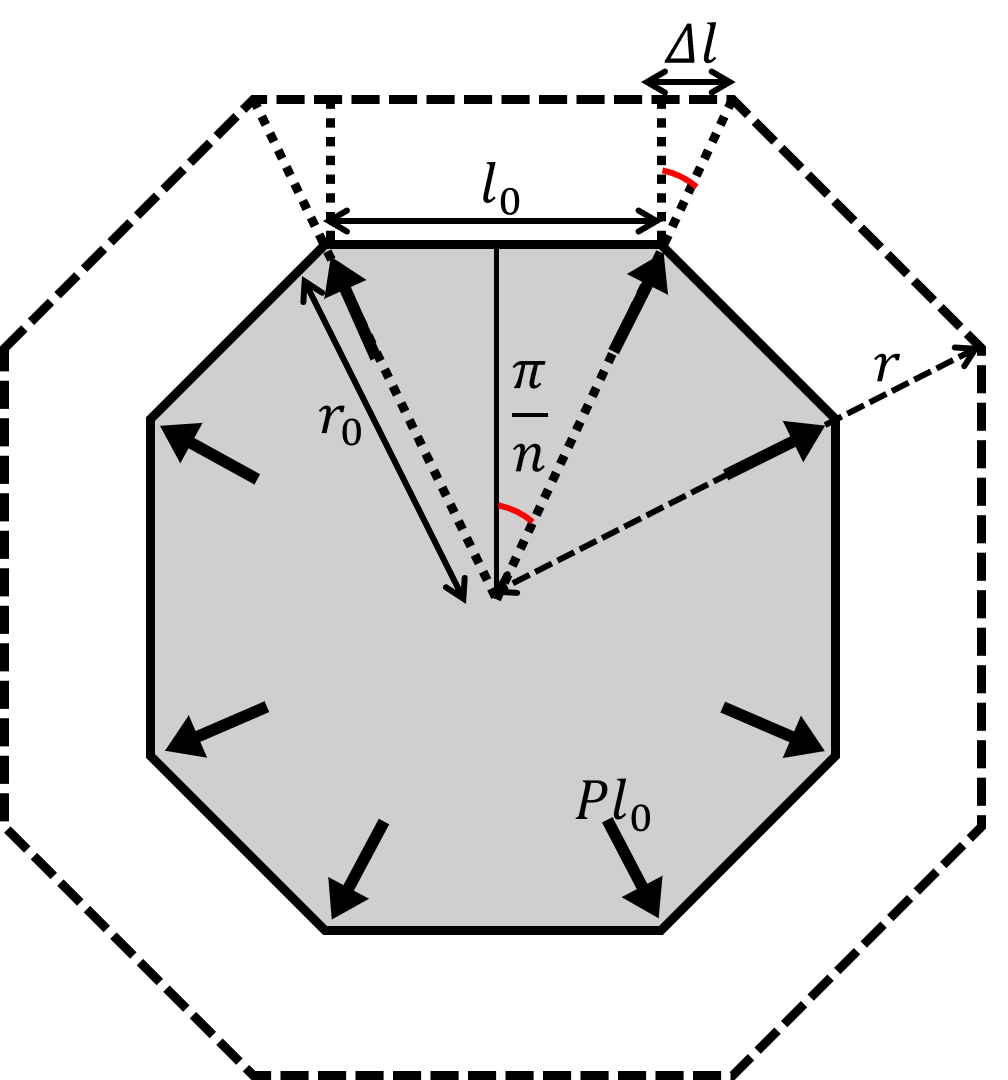} 
\caption{A single cell (at equilibrium) idealized as an $n$-sided regular polygon (here $n=8$). An isotropic pressure force $P l_0$ inflates the cell, pushing each vertex point outward and increasing the radius of the circumcircle from $r_0$ to $r$. Correspondingly, the surface springs at the edges extend from $l_0$ to ($l_0 + 2 \Delta l$). The diagram is not drawn to scale and is magnified for visual clarity. }
\label{fig1_schematic}
\end{figure}

\begin{justify}

The free energy of our model fundamentally differs from the topological energy function of Vertex/Voronoi-type models. For instance, consider a single isolated cell at equilibrium. Under an isotropic pressure, the cell will attain the equilibrium shape of an $n$-sided regular polygon having $n$ beads and $n$ springs (with spring constant $K_s$ and rest length $l_0$). Following Fig.~\ref{fig1_schematic},  the free energy is:
\begin{align}
F &= - n P l_0 (r - r_0) + \frac{n}{2} K_s (2 \Delta l)^2 \nonumber \\ 
       & = - n P l_0 (r - r_0) + 2 n K_s (r - r_0)^2 \text{sin}^2 (\pi/n),
\label{eq:comparison1}       
\end{align}
where the first term is the work done by the pressure, the second term is the spring energy, and we invoke the geometric relation, $\Delta l = (r - r_0) \text{sin}(\pi/n)$ (see Fig.~\ref{fig1_schematic}). Differentiating Eq.~({\ref{eq:comparison1}}) with respect to $r$, we obtain at equilibrium:
\begin{align}
\frac{\partial F}{\partial r} &= - n P l_0 + 4 n K_s (r - r_0) \text{sin}^2 (\pi/n) = 0.
\label{eq:comparison2}
\end{align}
Using the geometric relation, $l_0 = 2 r_0 \text{sin}(\pi/n)$ (see Fig.~\ref{fig1_schematic} ), Eq.~({\ref{eq:comparison2}}) can be reduced to
\begin{align}
P = K_s (2 \Delta l)/r_0 = T/r_0, 
\label{eq:comparison3}
\end{align}
where $T = K_s (2 \Delta l)$ can be considered as the local surface tension at each edge. The Eq.~({\ref{eq:comparison3}}) is basically the Young-Laplace (YL) equation, which is thus satisfied in our model. In contrast, the vertex energy function for a single cell is given by
\begin{align}
E &= \frac{K_P}{2} (p - p_0)^2 + \frac{K_A}{2} (a - a_0)^2, 
\label{eq:comparison4}
\end{align}
where $p$ and $a$ are perimeter and area of a cell, while $p_0$ and $a_0$ are target perimeter and target area parameters, respectively. The first term in Eq.~({\ref{eq:comparison4}}) is similar to the surface spring energy in our model. However, the 2nd term is the `area elasticity,' which represents a pressure dependent on the cell's instantaneous area (pressure, $P= K_A (a - a_0)$). This is inconsistent with the YL equation, which is arguably responsible for the tendency of rounding of a single cell during detachment \cite{cellrounding2023iscience, julicher2012science}. The YL equation was also invoked to quantify adhesion \cite{julicher2012science, heisenberg2021celltheo}, and it formed the basis of inferring tension at cellular edges from analysis of tissue images  \cite{shraiman2020prx, graner2021development}. Since the vertex energy function does not obey the YL equation, it can lead to skewed non-rounded shapes for a single cell at equilibrium, depending on independent choices of $a_0$ and $p_0$. Thus, a single isolated cell is poorly defined in this framework. An explicit `pressure' term must be incorporated into the energy function to make it consistent with the YL equation (see \cite{bubblyvertex2014pre}).\\
\end{justify}

\section{\label{sec:level4} MEASURED OBSERVABLES}
%\vspace{4cm}
\subsection{\label{sec:level4A} Mean Square Displacement (MSD)}
\begin{justify}
To analyze the cell motions in the monolayer, we measured the mean square displacement (MSD) of the cell centers. The MSD is defined as 
\begin{equation}
        \text{MSD}(\Delta{t}) :=  \langle{\Delta\textbf{\textit{r}}^2(\Delta{t})}\rangle %\nonumber \\
        %&= \Bigl\langle\overline{({\textbf{\textit{r}}_\alpha}^{cm}(t_0+\Delta{t}) - {\textbf{\textit{r}}_\alpha}^{cm}(t_0))^2}\Bigl\rangle % `\Bigl` is to enlarge the brackets `\langle`      
\end{equation}

where 
\begin{equation}
{\Delta\textbf{\textit{r}} = \textbf{\textit{r}}_\alpha}^{cm}(t_0+\Delta{t}) - {\textbf{\textit{r}}_\alpha}^{cm}(t_0) 
\label{eq:delta_r}
\end{equation}
${\textbf{\textit{r}}_\alpha}^{cm}(t) = (\sum_{i=1}^{M}\textbf{\textit{r}}_{i,\alpha})/M$ is the position vector of the center of mass of the $\alpha$-th cell at time instant $t$. The angular bracket $\langle{...}\rangle$ denotes the average over all the cells and over different ensemble runs.
\end{justify}

\subsection{\label{sec:level4B} Effective Diffusivity}
\begin{justify}
From the MSD, we determined the effective diffusivity $D_{\text{eff}}$ as a dynamical order parameter that differentiates the solid and fluid phases and is defined as 
\begin{align}
    D_{\text{eff}} :=  \lim_{t\to\infty}\frac{\text{MSD}(\Delta{t})}{4D_0\Delta{t}},
\end{align}
where $D_0 = {v_0}^2/2D_r$ is the self-diffusivity for an isolated single cell \cite{Bi2016}. $D_{\text{eff}}$ is estimated from the slope of the linear fits to the MSD curves at long time (from $5\times10^6$ to $10^7$ iterations).
\end{justify}

\subsection{\label{sec:level4C} Non-Gaussian Parameter}
\begin{justify}
The non-Gaussian parameter ($\alpha_2(\Delta{t})$), in 2D, is defined as \cite{Chiang2016} : 
\begin{align}
    \alpha_2(\Delta{t}) := \frac{1}{2}\frac{\langle{\Delta\textbf{\textit{r}}^4(\Delta{t})}\rangle}{{\langle{\Delta\textbf{\textit{r}}^2(\Delta{t})}\rangle}^2} - 1,
\end{align}
where $\Delta\textbf{\textit{r}}$ is as defined in Eq.~(\ref{eq:delta_r}). This parameter includes the $2^{nd}$ and $4^{th}$ order moments of the distribution of $\Delta\textbf{\textit{r}}$. The parameter value is zero for a Gaussian distribution and becomes non-zero when the distribution deviates from the Gaussian. The higher-order moments capture the deviation of the distribution, especially in the tail part, which cannot be accounted for by MSD.  
\end{justify}

\subsection{\label{sec:level4D} Area Fraction}
\begin{justify}
The area fraction in our model can be calculated in the following two ways. First, the area fraction ($\phi$) of the tissue monolayer at a particular time is given by
\begin{equation}
\phi = ({\sum_{\alpha=1}^{N}A_\alpha})/A_{box}
\label{eq:area_frac}
\end{equation}
Here $A_\alpha$ is the area of the $\alpha$-th cell, $N$ is the cell number, and $A_{box} = L^2$, is the area of the simulation box. This definition considers the beads as point particles. However, since we have a cut-off range of the repulsion interaction (${r_c}^{rep}$), each bead can be treated as a disc of diameter ${r_c}^{rep}$, contributing a factor of  $(\pi /2)({{r_c}^{rep}}/2)^2$ to the single cell area. Thus, according to this factor, we can have a higher value of the area fraction given by
\begin{equation}
\phi' = \phi + (M N/A_{box}) \pi {({r_c}^{rep})}^{2}/8
\label{eq:modified_area_frac}
\end{equation}
Here $M$ is the bead number per cell. We calculated that $\phi$ ranges $0.85-0.88$ across the extreme fluid and solid phases (corresponding to Fig. 2 of our manuscript), while $\phi' \sim 0.93-0.95$ during the transition. Both these estimates show high confluency, and the area fraction changes negligibly across the transition, as found during the liquid-to-solid transition in dense zebrafish tissues \cite{Mongera2018}.

\end{justify}

\subsection{\label{sec:level4E} Shape Index}
\begin{justify}
 
Following previous studies \cite{Bi2015,Bi2016}, 
 the cell shape index ($S$) is defined as: 
\begin{align}
    {S} = \frac{1}{N}\sum_{\alpha=1}^{N}\frac{P_\alpha}{\sqrt{A_\alpha}}
    \label{eq21_shapeind}
\end{align}
Where $P_\alpha$ is the perimeter and $A_\alpha$ is the area of the $\alpha$-th cell at a particular time instant. $S$ denotes the averaged shape index over all the cells at a particular instant. And, $\langle{S}\rangle$ indicates the average of $S$ over time and over different simulations (ensemble average). While calculating $\langle{S}\rangle$, we averaged over the time window from $5\times10^6$ to $10^7$ iterations when the system is deep in the steady state, as confirmed from the temporal evolution of shape index ($S$) (see Fig. 4E, main text).

We also measured $\langle{S}\rangle_{Vor}$, the averaged shape index obtained from the Voronoi tessellated polygons. Following standard protocol \cite{dp2018prl}, we performed the Voronoi space tessellation from the actual cell center data in a simulated tissue, to obtain the corresponding Voronoi polygons (see Fig.~\ref{fig:S9}A). Then we calculated the shape index, ${S}_{Vor}$, for each of the Voronoi polygons using Eq.~\ref{eq21_shapeind}. Note that, $\langle{S}\rangle_{Vor}$ represents the ensemble average as well as the time average in the steady-state (as stated above for measuring $\langle{S}\rangle$).

\end{justify}

\subsection{\label{sec:level4F} Edge Tension}
\begin{justify}
For the $i$-th bead of a particular cell at an instant, the edge tension is estimated as the extension of the corresponding spring length (apart from a multiplicative constant): 
\begin{equation}
    T_i = |l_i - l_0|
\end{equation}
where, $l_i$ is the instantaneous length of the spring between $i$-th and $(i-1)$-th beads of the cell and $l_0$ is the rest length. Then we calculated the average  tension over all the beads of a cell and  over all the cells as below:
\begin{equation}
    {T} = \frac{1}{N M}\sum_{\alpha=1}^{N}\sum_{i=1}^{M}T_i
\end{equation}  
\end{justify}

\subsection{\label{sec:level4G} Coordination Number}
\begin{justify}
 By definition, the number of cells with which a particular cell is `in contact', is the coordination number $(Z)$ of that particular cell \cite{2dfoam2017}. Now, in our model, a particular cell, say the $\alpha$-th cell, would be `in contact' with the $\beta$-th cell and vice versa, if any two beads of the two cells come within the interaction ranges, either ${r}^{rep}_c$ or ${r}^{adh}_c$. Thus, $Z$ of a particular cell at a particular time instant is the total number of cells that are interacting with it through either repulsion or adhesion. 
 \end{justify}

%\clearpage
%\newpage

%\newpage

\section{\label{sec:level5}Jamming onset in the `passive limit' of our model}
\begin{justify}    
 The `passive-tissue' limit of our model essentially means the absence of any active self-propulsion vector ($v_0=0$). In this limit, we studied the system without cell-cell adhesion ($K_{adh} = 0$) in order to compare with previous studies on the packing of soft grains \cite{Astrom2006pre, dp2018prl}. The intercellular interactions are purely repulsive in this case. We considered a system of $N=256$ cells in a square box of size $L = 460 l_0$, each cell containing $M = 50$ beads (same system size as described in the `Simulation Details' section). We followed the same initialization protocol stated in the `Simulation Details' section. Here, we performed the simulations using a `stiff' or `rigid' boundary condition. The stiff/rigid boundary condition was implemented as follows: At a particular time instant $t$, if any bead of a cell touches the boundary lines of the square box, i.e., if the Euclidean distance between the bead and any of the boundary lines becomes smaller than the range of repulsion ($r_c^{rep}$), then the bead position was not updated, and the bead was kept fixed to its previous position at the time instant $t-dt$ (where $dt$ is the integration timestep).

 The system was then evolved by integrating Eq.~\ref{eq:eom_position} with $\Tilde{K}_{adh} = 0$ and $\Tilde{v}_0 = 0$ (i.e., without Eq.~\ref{eq:eom_theta}) using the Euler method with integration time step $10^{-3}$. We equilibrated the system by iterating up to $10^5$ steps. The system then reached an equilibrium configuration with a particular level of confluency, depending on the value of the intracellular pressure $\Tilde{P}$. With increasing intracellular pressure, the equilibrium configurations slowly went from sub-confluent to highly confluent. Thus, we took a series of $\Tilde{P}$ values to generate configurations with increasing packing fractions (see Fig~\ref{fig:S8}(A-C)), similar to a protocol used in prior studies \cite{Astrom2006pre}.

 Finally, for each value of $\Tilde{P}$, we calculated the observable quantities like cell shape index, area fraction, coordination number, etc, 
by averaging over $50$ independent simulation runs. During the calculations of the quantities, we considered only the cells in the bulk, not those in contact with the box boundaries. 
Notably, we observed the onset of crowding-dependent jamming in our passive model similar to the Deformable Particle (DP) model \cite{dp2018prl} published before (see Fig.~\ref{fig:S8}(D-E)). As shown in Fig.~\ref{fig:S8}D, the tissue reaches the confluence when the normalized area fraction reaches unity ($\phi/\phi_{max} = 1.0$ with $\phi_{max} = 0.883$), at the measured shape index $\langle{S}\rangle \approx 3.81$. The average coordination number $Z \sim 5.7$ indicates a jammed confluent tissue at $\langle{S}\rangle = 3.81$ (Fig.~\ref{fig:S8}E). Note that this value of $Z$ at the confluence is higher than the isostatic contact number for the Vertex model  ($Z_{iso}=5$), as argued from the Voronoi tesselation of the 2D space \cite{Bi2015}.  Interestingly, the Self-propelled Voronoi (SPV) model \cite{Bi2016} also shows a liquid-to-solid transition at $\langle{S}\rangle \approx 3.81$ in the zero motility limit. Also note that our model, despite having a distinct single-cell energy function (in equilibrium) compared to the DP and Vertex models, still shows the jamming onset at $\langle{S}\rangle \approx 3.81$. This indicates that the jamming onset can be explained from the argument of isostaticity based on the polygonal tiling of the 2D space (since $Z_{iso}=5$ corresponds to the $\langle{S}\rangle = 3.81$ for a pentagon), as described in \cite{Bi2015}.

\end{justify}

\section{\label{sec:level6}The confluent limit of our model}

\begin{justify}
    
    As discussed in the previous section (Section V of the SI), the level of confluency may be described by the average area fraction normalized by the maximum area fraction in a parameter regime ($\phi/\phi_{max}$). The tissue becomes confluent as $\phi/\phi_{max} \to 1$. Note that a particle-based description like our model cannot have $\phi_{max}=1$ since two adjacent cells have distinct, non-overlapping adjoining edges unlike the Voronoi-tessellation-based models, where adjacent cells share a common edge with `zero-thickness'. In our case, $\phi_{max} \to 1$ only when the bead number per cell approaches infinity. Nevertheless, 100\% confluency is an idealization since {\it in vivo} tissues still show 80\%-90\% confluency (see Fig. 2e, 2h in \cite{Mongera2018}).

    In the main paper, we kept the tissue at a near-confluent regime (where $(\phi/\phi_{max}) \sim 0.94-1$ across the transition). We calculated that $\phi$ ranges $0.83-0.88$ across the fluid and solid phases (corresponding to Fig. 2-4, main text), while $\phi' \sim 0.91-0.96$ during the transition (note the slightly different definition of packing fractions, $\phi$, and $\phi'$, as given in Eq. \ref{eq:area_frac}-\ref{eq:modified_area_frac} in SI). These estimates show near-confluency, and the area fraction hardly changes across the transition.

As mentioned above, since our model does not attain $\phi=1$, we cannot directly compare the average shape indices of cells with the shape indices of the Voronoi cells in the Vertex and SPV models. Thus, for comparison, we performed the Voronoi tessellation of space (using the cell centers) to obtain the Voronoi polygons corresponding to the cells (Fig.~\ref{fig:S9}A), and we then measured the shape indices of the Voronoi cells ($\langle{S}\rangle_{Vor}$, as defined in Sec.~\ref{sec:level4E}, SI). Interestingly, in this near-confluent regime, $\langle{S}\rangle_{Vor}$ measured from the Voronoi polygons shows that the liquid-solid transition takes place at $\langle{S}\rangle_{Vor} \approx 3.81$ for the low values of rotational noise strength ($D_r$), similar to the behavior of Vertex and SPV models \cite{Bi2015, Bi2016} (see Fig.~\ref{fig:S9}B,C). However, for higher $D_r$, the phase boundary shifts from $\langle S \rangle_{Vor} = 3.81$ (Fig.\;\ref{fig:S9}C). Thus, our model replicates the behavior of the Vertex model in certain regimes if we treat $\langle{S}\rangle_{Vor}$ as a structural order parameter. But, note that the actual shape index $\langle{S}\rangle$ measured from cell contours are nonmonotonic near the phase boundary (Fig. 4F, main text), contrasting the behavior of the Vertex model.

Furthermore, as our model has tunable confluency depending on parameter variations, we could achieve high confluency by increasing the natural spring length ($l_0$ from 0.1 to 0.14) and explored the effect of increasing adhesion from zero to a much higher value (see Fig.~\ref{fig:S10}(A-B)). Though the packing fraction did not alter up to the second decimal place ($\phi = 0.891 - 0.893$), the MSD of cell centers still showed a liquid-solid transition, where the cell dynamics changed from sub-diffusive to caged (plateau) behavior with adhesion (Fig.~\ref{fig:S10}C, D). The transition points are located based on a cut-off value of the MSD exponent (see Fig.~\ref{fig:S10}D, where the cut-off exponent was set to 0.2). Notably, the average packing fraction (both $\phi$ and $\phi'$)  hardly changed across the liquid-solid transition, suggesting a rigidity-driven glass transition, not a crowding-dependent jamming (Fig.~\ref{fig:S10}E). Also, note that the values of the area fraction in Fig.~\ref{fig:S10}E are well above the critical packing fraction for jamming transition in passive foams --- in previous simulations of the `deformable polygon model' and `dynamic vertex model,' the critical packing fraction for jamming was reported to be $\phi_c \sim 0.83$ in 2D (see Fig. 2f in \cite{Kim2021} and Fig. 4a in \cite{dp2019softmatt}). Thus, for the chosen parameters of our model, the tissue is already `jammed' without adhesion, showing sub-diffusive dynamics with active cellular motility (Fig.~\ref{fig:S10}C). However, with increasing adhesion, solidification arises from dampening the active motility of cells, which is caused by the increased resistance to shear forces between cells due to higher adhesion.

Moreover, the solidification in the highly confluent state accompanied a negligible change in the average contact number per cell ($Z$ changes from 5.90 to 5.98 in Fig.~\ref{fig:S10}F corresponding to the red and blue curves, representing the confluent limit). In this regard, a simple counting of constraints for passive non-adhesive spheres shows that the critical isotactic contact number for jamming is $Z_c \sim 4$ in 2D. Note, the degrees of freedom for $N$ cells is $n_{DOF} = 2N$, while the number of constraints is $n_{c} = Z N/2$ as each contact is shared between two cells. This leads to $Z_c =4$ when $n_{DOF}=n_{c}$ (see \citep{manning2021review}). However, another constraint counting argument based on the Voronoi tesselation of the 2D space suggests that the isostatic value of $Z$ at the confluence should be $Z_{iso}^{vor}=5$, as explained for the Vertex model \cite{Bi2015}. Importantly, as shown in Fig.~\ref{fig:S10}F, our observed contact number in the confluent limit is above both the estimated isostatic values ($Z_{c}=4$ and $Z_{iso}^{vor}=5$). Thus, the solidification is essentially an adhesion-dependent active glass transition, not crowding-dependent jamming.

 Additionally, in the confluent limit, we measured both the average shape indices from the Voronoi cells ($\langle{S}\rangle_{Vor}$) and from the actual cell contours ($\langle{S}\rangle$) (Fig.~\ref{fig:S10}G, H). Notably, the liquid-solid transition boundaries (pinpointed by arrowheads in Fig.~\ref{fig:S10}D) closely coincide with $\langle{S}\rangle_{Vor} \approx 3.81$  (Fig.~\ref{fig:S10}G), again exhibiting the Vertex and SPV model-like behavior \cite{Bi2015, Bi2016} for the given parameter choices. Nevertheless, the actual shape indices still show non-monotonic behavior across the transition (Fig.~\ref{fig:S10}H), similar to what was observed before (as in Fig. 4F, main text). 
   
\end{justify}

\newpage

\section{\label{sec:level7} SUPPLEMENTAL FIGURES}

%\clearpage
%\newpage

%\vspace{4cm}

\begin{figure*}[hbt!]
\includegraphics[width=12cm]{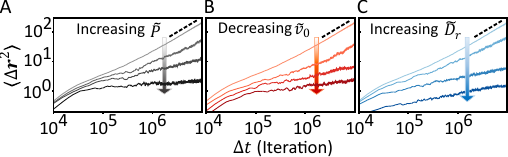}% Here is how to import EPS art
\caption{ \textbf{Fluid to solid transition with respect to three dimensionless parameters.} The mean-squared displacements of cell centers show transitions from the fluid-like diffusive to the solid-like sub-diffusive behavior \textbf{(A)} with increasing intracellular pressure (top to bottom: $\tilde{P} = 0.20, 0.22, 0.24, 0.33$), \textbf{(B)} with decreasing motility (top to bottom: $\tilde{v}_0 = 16.6 \times10^{-3}, 15.0 \times10^{-3}, 13.2 \times10^{-3}, 10.8 \times10^{-3}$), and \textbf{(C)} with increasing rotational noise strength (top to bottom: $\tilde{D}_r = 5.2 \times10^{-4}, 7.5 \times10^{-4}, 13.0 \times10^{-4}, 47 \times10^{-4}$). Parameters: $\tilde{v}_0 = 16.6\times10^{-3}$, $\tilde{P} = 0.2$, $\tilde{D}_r = 5.2\times10^{-4}$, $\tilde{K}_{adh} = 8.3\times10^{-4}$ (Note that one of these parameters are varied in panels A-C). Other parameters are from Table S1.}
\label{fig:S3}
\end{figure*}

\begin{figure*}[hbt!]\includegraphics[width=12.5cm]{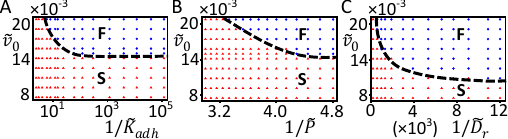}% Here is how to import EPS art
\caption{ \textbf{Phase diagrams in the parameter space of (A)}  $\tilde{v}_0$ vs. $1/\tilde{K}_{adh}$, \textbf{(B)} $\tilde{v}_0$ vs.  $1/\tilde{P}$, and \textbf{(C)} $\tilde{v}_0$ vs. $1/\tilde{D}_r$. Dashed lines correspond to the phase boundaries. Parameters: $\tilde{P} = 0.2$, $\tilde{D}_r = 5.2\times10^{-4}$, $\tilde{K}_{adh} = 8.3\times10^{-4}$ (Note that one of these parameters are varied along with $\tilde{v}_0$ in panels A-C). Other parameters are from Table S1.}
\label{fig:S4}
\end{figure*}

\begin{figure*}[hbt!]
\includegraphics [width=9.6cm]{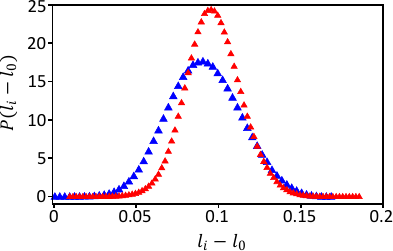}% Here is how to import EPS art
\caption{\textbf{Edge-tension distributions.} Probability distributions of the edge elongation $(l_i-l_0)$ in the tissue kept in fluid (blue) and solid (red) phases, corresponding to Fig. 4C (main paper). 
 Parameters: $\tilde{K}_{adh} = 8.3\times10^{-6}$, $\tilde{D}_r = 0.8\times10^{-4}$ (blue) and $\tilde{K}_{adh} = 0.42$, $\tilde{D}_r = 0.001$ for B. Other parameter values are as mentioned in Fig. 3 caption.}
\label{fig:S5}
\end{figure*}

\begin{figure*}[hbt!]
\includegraphics [width=9.6cm]{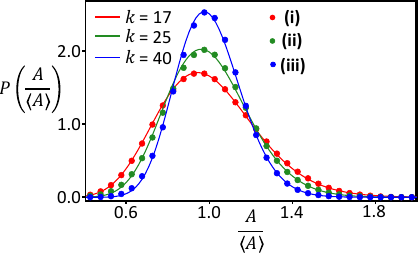}% Here is how to import EPS art
\caption{\textbf{Cell-area distributions.} Probability distributions of the scaled cell-area, $\tilde{A} = A/\langle{A}\rangle$, for region-i (red), ii (green), and iii (blue), corresponding to Fig. 4F. Solid lines are the fits with the generalized $k$-Gamma distribution, $\frac{k^k}{\Gamma(k)}\tilde{A}^{k-1}exp(-k\tilde{A})$. Note that the $k$ value increases with an increase in $\tilde{K}_{adh}$ as the tissue progressively solidifies. Here, $\langle{A}\rangle$ denotes the mean of cell-area. Parameters: Same as Fig. 4F.}
\label{fig:S6}
\end{figure*}

\begin{figure*}[hbt!]\includegraphics*{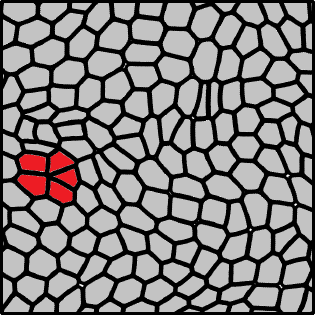}% Here is how to import EPS art
\caption{\textbf{Spontaneous formation of a muticellular rosette.} A configuration showing a five-cell junction (highlighted in red) known as a rosette that can be formed when five or more cells share a vertex. Parameters: $\tilde{K}_{adh} = 8.3\times10^{-5}$, $\tilde{P} = 0.2$, $\tilde{v}_0 = 16.6\times10^{-3}$, $\tilde{D}_r = 5.2\times10^{-4}$, $l_0=0.14$. Other parameters are from Table S1.}
\label{fig:S7}
\end{figure*}

\begin{figure*}[hbt!]\includegraphics*[width=14cm]{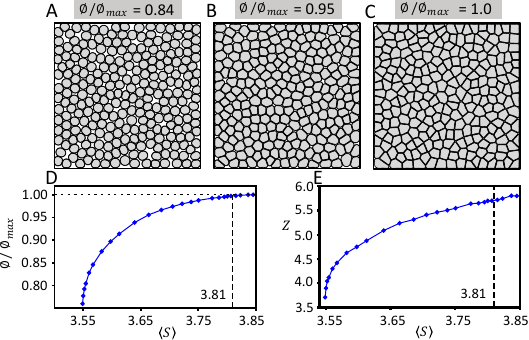}% Here is how to import EPS art
\caption{\textbf{Jamming onset in the `passive limit'.} 
We obtained the `passive limit' of our model by switching off the self-propulsion speed ($v_0 = 0$) in the absence of cell-cell adhesion ($K_{adh} = 0$). Jammed states are formed by systematically increasing the intracellular pressure (consequently increasing the packing fraction) in a box with rigid boundaries. For more details, see the SI text, Section V. \textbf{(A-C)} Steady state simulation snapshots with increasing packing fractions: $\phi/{\phi}_{max} = 0.84$ (A), $0.95$ (B), and $1.0$ (C). \textbf{(D)} Normalized area fraction $(\phi/{\phi}_{max})$ and \textbf{(E)} the average contact number per cell ($Z$), as functions of the averaged shape index $(\langle{S}\rangle)$. The tissue reaches the confluence when the normalized packing fraction becomes unity ($\phi/{\phi}_{max} = 1$) at the measured shape index $\langle{S}\rangle \approx 3.81$ (shown by the vertical dashed line in D). The maximal packing fraction achieved is $\phi_{max} = 0.883$, as shown by the horizontal dotted line in panel D.
 The average coordination number $Z \sim 5.7$ indicates a jammed confluent state at $\langle{S}\rangle = 3.81$ (indicated by the vertical dashed line in E). Parameters: $\tilde{K}_{adh} = 0$, $v_0 = 0$, $l_0 = 0.1$ and varying $\tilde{P}$ from 0.0875 to 0.75. Other parameters are from Table S1.} 
\label{fig:S8}
\end{figure*}

\begin{figure*}[hbt!]\includegraphics*[width=15cm]{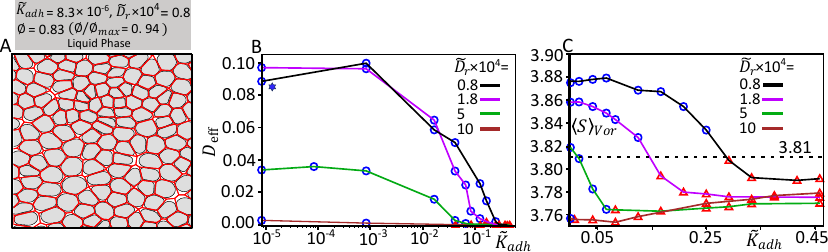}% Here is how to import EPS art
\caption{\textbf{Calculation of shape indices by Voronoi tessellation (corresponding to Fig. 4F).} \textbf{(A)} A configuration of polygonal cells (red solid lines), encompassing the original cells (grey), obtained from the Voronoi tesselation of the space in the liquid phase (corresponding to the blue star-marked point in panel B). \textbf{(B)} The effective diffusivity from long time MSD slope, $D_\texttt{eff}$, is shown against $\tilde{K}_{adh}$ for different values of $\tilde{D}_r$ (rotational noise strength). Data points with $D_\texttt{eff} > 0.001$ are marked as liquid (blue circles), and $D_\text{eff} \leq 0.001$ are marked as solid (red triangles). \textbf{(C)} The average shape index measured from the Voronoi polygons($\langle S \rangle_{Vor}$) is plotted with adhesion strength ($\tilde{K}_{adh}$) for different values of $\tilde{D}_r$. The horizontal dashed line corresponds to $\langle{S}\rangle_{Vor} = 3.81$. The black, green, and brown curves in panel B correspond to Fig. 4F of the main paper. In Panels B and C, blue circles and red triangles represent fluid and solid phases, respectively. Parameters are from Fig. 4 of our manuscript and Table S1.}
\label{fig:S9}
\end{figure*}

\begin{figure*}[hbt!]\includegraphics*[width=15cm]{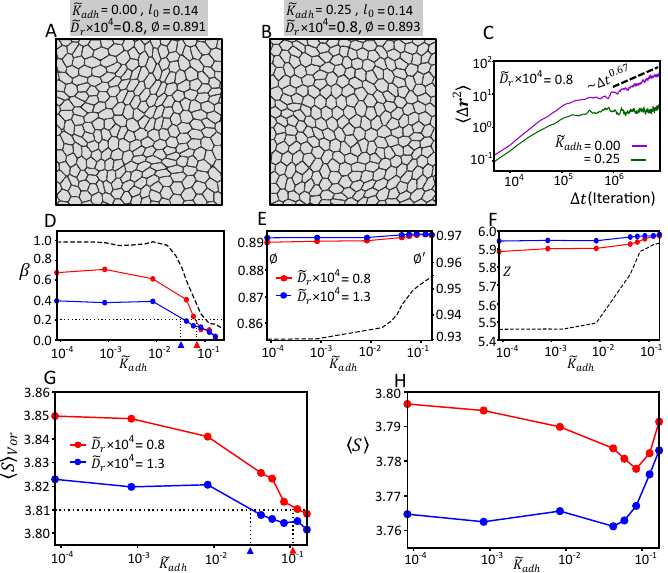} 
\caption{\label{fig:epsart} \textbf{Solidification at highly confluent tissues.} 
\textbf{(A-B)} Tissue configurations for $\tilde{K}_{adh} = 0$ (A) and $\tilde{K}_{adh} = 0.25$ (B). Note that high confluence was achieved by increasing $l_0$. \textbf{(C)} MSD of cell centers (corresponding to $\tilde{K}_{adh}$ values as in panels A and B), showing a transition from sub-diffusive to caged behavior. The dashed line shows the MSD exponent of $0.67$ for $\tilde{K}_{adh} = 0$. \textbf{(D)} MSD exponent ($\beta$) shows transition with $\tilde{K}_{adh}$ at a lower $\tilde{D}_r$ (red curve) and at a higher $\tilde{D}_r$ (blue curve). The blue and red arrowheads indicate the transition points (and corresponding $\tilde{K}_{adh}$ values) based on a cut-off, $\beta=0.2$, as shown by the dotted horizontal line. \textbf{(E)} The area fraction (both $\phi$ and $\phi'$, as defined in Sec. IVD, SI) and \textbf{(F)} the average contact number per cell ($Z$), are plotted against $\tilde{K}_{adh}$ for the highly confluent tissues at two different $\tilde{D}_r$ values (red and blue ones). Note that in panels D, E, and F, the black dashed curve is for the parameters corresponding to Fig. 2E (main text), given here for comparison with a near-confluent tissue. Notably, $\phi$ and $Z$ are almost constant for the fully confluent tissue compared to the near-confluent tissue denoted by the dashed reference curves. 
\textbf{(G)} The average shape index measured from the Voronoi tessellated polygons ($\langle{S}\rangle_{Vor}$), and \textbf{(H)} the average shape index ($\langle{S}\rangle$) obtained from the actual cell contours, are plotted against $\tilde{K}_{adh}$. In G, the blue and red arrowheads indicate the $\tilde{K}_{adh}$ values corresponding to $\langle{S}\rangle_{Vor} = 3.81$, which coincide with the liquid-solid transition points located in panel D. For all panels, $l_0 = 0.14$ except the dashed reference curve (in D-F), for which $l_0=0.1$.  Other parameters are from Fig. 3 (main text) and Table S1.}
\label{fig:S10}
\end{figure*}

\clearpage
\newpage

\section{\label{sec:level8} SUPPLEMENTAL MOVIE CAPTIONS}
\begin{justify}
%The following are the captions regarding the supplemental movies. \\

 Movie S1. \textbf{Fluid tissue phase (low adhesion regime)}. The steady-state dynamics of the tissue exhibiting a liquid-like phase at (low)  adhesion strength $\Tilde{K}_{adh} = 8.3 \times10^{-5}$. A group of neighboring cells are marked in red and green to highlight a T1 transition. Also, note the freely diffusive dynamics of the cells. Parameter values: $\Tilde{P} = 0.2$, $\Tilde{v}_0 = 16.6\times10^{-3}$, $\Tilde{D}_r = 5.2\times10^{-4}$. Other parameters are from Table S1.\\

 Movie S2. \textbf{Solid tissue phase (high adhesion regime)}. The steady-state dynamics of the tissue exhibiting a solid-like phase at (high) adhesion strength, $\Tilde{K}_{adh} = 0.25$. A randomly chosen cell is marked red to highlight that its motion is completely caged by its neighbors. Rest of the parameter values are the same as in Movie S1. 
\end{justify}

\bibliography{ref.bib}
%\appendix

%\section{Appendixes}

%\nocite{*}


\end{document}